\documentclass[times,twocolumn,final]{elsarticle}
\usepackage{medima}
\usepackage{framed,multirow}

\usepackage{amssymb}
\usepackage{latexsym}

\usepackage{url}
\usepackage{xcolor}

\usepackage{hyperref}

\definecolor{newcolor}{rgb}{.8,.349,.1}

\usepackage{booktabs}
\usepackage{multirow}
\usepackage{pifont}
\newcommand{\cmark}{\ding{51}}%

\journal{Medical Image Analysis}

\begin{document}

% \verso{Given-name Surname \textit{et~al.}}
\verso{Qiang Ma \textit{et~al.}}

\begin{frontmatter}

\title{The Developing Human Connectome Project: A Fast Deep Learning-based Pipeline for Neonatal Cortical Surface Reconstruction}

% \tnoteref{tnote1}}
% \tnotetext[tnote1]{This is an example for title footnote coding.}

\author[1]{Qiang Ma\corref{cor1}}
\cortext[cor1]{Corresponding author.}
\ead{q.ma20@imperial.ac.uk}
\author[2,3]{Kaili Liang}
%\fnref{fn1}}
% \fntext[fn1]{This is author footnote for second author.}
\author[1]{Liu Li}
%% Third author's email
% \ead{author3@author.com}
\author[2]{Saga Masui}
\author[2]{Yourong Guo}
\author[2,3]{Chiara Nosarti}
\author[2]{Emma C. Robinson}
\author[1,4]{Bernhard Kainz}
\author[1,5]{Daniel Rueckert}

\address[1]{Department of Computing, Imperial College London, UK}
\address[2]{School of Biomedical Engineering and Imaging Sciences, King's College London, UK}
\address[3]{Department of Child and Adolescent Psychiatry, Institute of Psychiatry, Psychology and Neuroscience, King's College London, UK}
\address[4]{FAU Erlangen–N\"urnberg, Germany}
\address[5]{Chair for AI in Healthcare and Medicine, Klinikum rechts der Isar, Technical University of Munich, Germany}

\received{14 May 2024}
% \finalform{10 May 2013}
\accepted{12 Nov 2024}
% \availableonline{15 May 2013}
% \communicated{S. Sarkar}

\begin{abstract}
%%%

The Developing Human Connectome Project (dHCP) aims to explore developmental patterns of the human brain during the perinatal period. An automated processing pipeline has been developed to extract high-quality cortical surfaces from structural brain magnetic resonance (MR) images for the dHCP neonatal dataset. However, the current implementation of the pipeline requires more than 6.5 hours to process a single MRI scan, making it expensive for large-scale neuroimaging studies. In this paper, we propose a fast deep learning (DL) based pipeline for dHCP neonatal cortical surface reconstruction, incorporating DL-based brain extraction, cortical surface reconstruction and spherical projection, as well as GPU-accelerated cortical surface inflation and cortical feature estimation. We introduce a multiscale deformation network to learn diffeomorphic cortical surface reconstruction end-to-end from T2-weighted brain MRI. A fast unsupervised spherical mapping approach is integrated to minimize metric distortions between cortical surfaces and projected spheres. The entire workflow of our DL-based dHCP pipeline completes within only 24 seconds on a modern GPU, which is nearly 1000 times faster than the original dHCP pipeline. The qualitative assessment demonstrates that for 82.5\% of the test samples, the cortical surfaces reconstructed by our DL-based pipeline achieve superior (54.2\%) or equal (28.3\%) surface quality compared to the original dHCP pipeline. 

%%%%
\end{abstract}


\end{frontmatter}

%\linenumbers

\section{Introduction}
\subsection{Background}
During the perinatal phase, the human brain undergoes significant developmental changes, characterized by the formation of anatomical and functional connectivity. The Developing Human Connectome Project (dHCP) plays a pivotal role in this context. It aims at compiling an extensive dataset of brain magnetic resonance imaging (MRI) for both fetuses and neonates \citep{hughes2017dedicated,makropoulos2018dhcp,edwards2022release}. This ambitious project is instrumental in facilitating an in-depth understanding of early brain development and in constructing a four-dimensional connectome for young brains. The dHCP dataset is a valuable asset for investigating the normal and abnormal patterns of brain structural development and their respective connectomes. Notably, the dHCP has publicly released its neonatal dataset, comprising MRI scans of 783 newborn infants, both pre-term and full-term, resulting in a total of 887 images \citep{edwards2022release}.

Cortical surfaces refer to the inner and outer surfaces of the cerebral cortex. The inner surface, also called white matter surface, is the boundary between the white matter (WM) and cortical grey matter (cGM). The outer surface, also called pial surface, is the boundary between the cGM and cerebrospinal fluid (CSF). Cortical surfaces are often represented by 3D polygon meshes and should have genus-0 topology, \emph{i.e.}, topologically equivalent to a 2-sphere without any "holes". 
The cortical folding and brain development of newborn infants can be modeled and quantified by the morphological features of cortical surfaces such as cortical thickness, mean curvature and sulcal depth \citep{nie2012growth,garcia2018folding}. The notable achievements of FreeSurfer \citep{dale1999cortical,fischl1999cortical,fischl2012freesurfer} and the Human Connectome Project (HCP) \citep{van2013hcp,glasser2013hcp}, recognized for their automated neuroimaging processing pipelines for adults, underscore the importance of implementing cortical surface-based structural brain MRI processing and analysis for the dHCP fetal and neonatal datasets.

However, the pronounced differences between adult and neonatal brains present significant challenges in adapting existing adult MRI processing pipelines for use in newborn infants \citep{fischl2012freesurfer,glasser2013hcp,shattuck2002brainsuite,macdonald2000civet,kim2005civet}. One primary challenge lies in the limited resolution of neonatal brain MRI, a consequence of the necessarily brief data acquisition times \citep{hughes2017dedicated}. This limitation is further compounded by the smaller head size in neonates, which results in a reduced region of interest (ROI) \citep{orasanu2014volume,makropoulos2016regional}. Additionally, the rapid developmental changes in the neonatal brain contribute to a significant variability in the geometric shape and size of the brain across different ages. Another notable aspect is the reversed contrast in neonatal brain MRI, attributed to the ongoing myelination process in the white matter, which is not yet complete \citep{prastawa2005contrast}. As a result of these factors, T2-weighted (T2w) images are predominantly used in the structural MRI processing of neonatal brains.

\subsection{Related Work}
To address the challenges of neonatal cortical surface extraction, \cite{makropoulos2018dhcp} proposed an automated structural brain MRI processing pipeline for dHCP neonatal dataset. The pipeline consists of brain extraction, bias field correction, brain segmentation, cortical surface reconstruction, inflation and parcellation, cortical feature estimation, and spherical projection. The dHCP neonatal structural pipeline \citep{makropoulos2018dhcp} employs the FMRIB software library (FSL) brain extraction tool (BET) \citep{smith2002bet,jenkinson2012fsl} for skull stripping and the N4 algorithm \citep{tustison2010n4} for bias correction. The Draw-EM approach \citep{makropoulos2014drawem} is adopted for brain regional and tissue segmentation. A deformable model-based approach \citep{schuh2017deformable} is introduced for cortical surface reconstruction. Such an approach iteratively deforms an initial convex hull surface to fit the zero level set of an implicit surface defined by the WM segmentation. Based on the T2w MRI intensity, the surface is then deformed towards the WM/cGM and cGM/CSF interface to generate white matter and pial surfaces. The spherical topology is guaranteed and no topology correction \citep{segonne2007topology,bazin2005topology,bazin2007topology} is required. For spherical projection, the dHCP pipeline adopts the Spherical Multi-Dimensional Scaling (MDS) approach \citep{elad2005mds} to minimize the geodesic distance distortions between the sphere and the white matter surface. The remaining parts of the pipeline follow the design of FreeSurfer \citep{fischl2012freesurfer} and the HCP pipeline \citep{glasser2013hcp}.

The dHCP structural pipeline has shown its advantages on cortical surface-based neonatal MRI analysis \citep{makropoulos2018dhcp} and provides crucial support for downstream applications \citep{schuh2018atlas,bozek2018atlas,fetit2020fetalseg}. However, the current dHCP pipeline, as well as other conventional infant neuroimage processing pipelines such as infant FreeSurfer \citep{zollei2020infant} and iBEAT \citep{dai2013ibeat}, still have limitations. First of all, these pipelines require several hours to process a single brain MRI scan, which are computationally expensive for large-scale neuroimage studies. Besides, the quality of the cortical surfaces extracted by the pipelines heavily relies on the brain tissue segmentations. Therefore, inaccurate segmentations can cause subsequent corruptions in the cortical surfaces, especially in the parietal and occipital lobes where the T2w MRI have low image contrast \citep{makropoulos2014drawem}.

To overcome the limitations in the traditional neuroimage pipelines, one alternative solution is deep learning (DL), which trains a deep neural network on a large dataset and makes end-to-end predictions during inference. DL-based approaches are supported by mature frameworks such as TensorFlow \citep{abadi2016tensorflow} and PyTorch \citep{paszke2019pytorch}, and accelerated by modern graphics processing units (GPUs) with fast parallel computation. Recently, several DL-based brain MRI processing pipelines have been proposed such as FastSurfer \citep{henschel2020fastsurfer}, DeepCSR \citep{cruz2021deepcsr}, SegRecon \citep{gopinath2021segrecon,gopinath2023segrecon}, and iBEAT V2.0 \citep{wang2023ibeat}. These pipelines introduce learning-based modules to predict implicit surfaces from brain MRI for cortical surface reconstruction. The implicit surfaces are represented as a segmentation or a signed distance function (SDF), which is a level set indicating the signed distance to the surface \citep{park2019deepsdf}. An explicit 3D mesh can be extracted from the implicit surface using the Marching Cubes algorithm \citep{lorensen1998mc}. Subsequently, topology correction algorithms \citep{segonne2007topology,bazin2005topology,bazin2007topology} are usually applied to fix topological errors in the extracted surface such that it is topologically equivalent to a sphere. However, such topology correction algorithms are usually time-consuming as they either repeatedly project the surface onto a sphere to detect topological defects \cite{segonne2007topology}, or iterate over all voxels to detect critical points \citep{bazin2005topology,bazin2007topology}. Hence, current DL-based pipelines \citep{henschel2020fastsurfer,wang2023ibeat,billot2023seg} still require more than half an hour to process a single subject, although they can extract accurate cortical surfaces.

In contrast to learning implicit surfaces,  explicit learning-based approaches \citep{wickramasinghe2020voxel2mesh,hoopes2021topofit,lebrat2021corticalflow,santa2022corticalflow,ma2021pialnn,ma2022cortexode,ma2023cotan,bongratz2022vox2cortex,bongratz2024v2cflow,zheng2024coupled} predict a sequence of deformations from brain MRI to deform an initial template surface explicitly into cortical surfaces. Since explicit deformations do not change the surface topology, time-consuming topology correction is not necessary if the initial surface has genus-0 topology. Therefore, explicit approaches, being both fast and capable of end-to-end learning, require only a few seconds to extract cortical surfaces, effectively circumventing the corruptions caused by imprecise segmentations. 
{In order to learn high-quality manifold meshes \citep{gupta2020nmf}, diffeomorphic deformations, which are widely used in medical image registration \citep{beg2005lddmm,ashburner2007dartel,vercauteren2009demon,balakrishnan2019voxelmorph}, have been incorporated into recent explicit learning-based cortical surface reconstruction approaches} such as CortexODE \citep{ma2022cortexode}, CorticalFlow \citep{lebrat2021corticalflow}, CorticalFlow++ \citep{santa2022corticalflow}, CoTAN \citep{ma2023cotan}, V2C-Flow \citep{bongratz2024v2cflow}, and \cite{zheng2024coupled}. Such approaches aim to learn diffeomorphic surface deformations to preserve the surface topology and prevent surface self-intersections, which will affect the estimation of cortical geometric features such as cortical volume, surface area and cortical thickness \citep{ma2022cortexode}.

After surface reconstruction, spherical projection or spherical mapping, is usually performed to project the cortical surface onto a spherical mesh while minimizing their metric distortions \citep{fischl1999cortical}, \emph{e.g.}, the distortions of edge length, face area, or geodesic distance. The spherical projection allows the cortical surface to be represented in a spherical coordinate system for downstream tasks such as cortical surface registration and atlas construction \citep{fischl1999sphere,robinson2014msm,robinson2018msm,bozek2018atlas}. Conventional spherical projection approaches \citep{fischl1999cortical,elad2005mds} are computationally intensive due to the iterative inflation of cortical surfaces and optimization of metric distortions. Recently, \cite{zhao2022sphereproj} proposed an end-to-end unsupervised learning framework for spherical mapping. Multiscale Spherical U-Net models \citep{zhao2019sphericalunet, zhao2019parcellation} are employed to learn coarse-to-fine diffeomorphic deformation fields to deform an initial projected sphere. The Spherical U-Nets are trained by minimizing the unsupervised metric distortions between the deformed spheres and resampled cortical surfaces for each resolution.

\subsection{Contributions}
Based on the original dHCP structural pipeline \citep{makropoulos2018dhcp}, we propose a new DL-based pipeline for dHCP neonatal cortical surface reconstruction. Such a pipeline comprises brain extraction, bias field correction, cortical surface reconstruction, cortical surface inflation, cortical feature estimation, and spherical projection. A brain mask is learned by a 3D U-Net \citep{ronneberger2015unet} for skull stripping. A multiscale diffeomorphic deformation network is proposed to extract cortical surfaces from the T2w MRI end-to-end without the requirement of tissue segmentations. We refine the original unsupervised spherical mapping framework \citep{zhao2022sphereproj} such that the runtime is accelerated from 2.1 to 0.3 seconds, while the edge and area distortions are reduced significantly. Following the procedure of the original dHCP pipeline, the remaining steps of the pipeline are re-implemented with GPU acceleration.

Our DL-based dHCP neonatal pipeline is primarily implemented in Python. Most of the processing steps in the pipeline are implemented with PyTorch \citep{paszke2019pytorch}, a well-known deep learning library enabling GPU acceleration. Our DL-based pipeline is fast and memory-efficient, which only needs 24 seconds to complete all procedures on a modern GPU with 12GB memory (or less than 3 minutes on CPUs only). This is nearly 1000$\times$ faster than the original dHCP pipeline, which requires more than 6.5 hours of runtime. We perform manual quality control of the cortical surfaces in the test set, which demonstrates that our dHCP DL-based pipeline produces superior cortical surface quality than the original dHCP pipeline \citep{makropoulos2018dhcp} while being orders of magnitude faster. The code for our DL-based dHCP neonatal pipeline is publicly available\footnote{\url{https://github.com/m-qiang/dhcp-dl-neonatal/}}.

\begin{figure}[ht]
\centering
\includegraphics[width=1.0\linewidth]{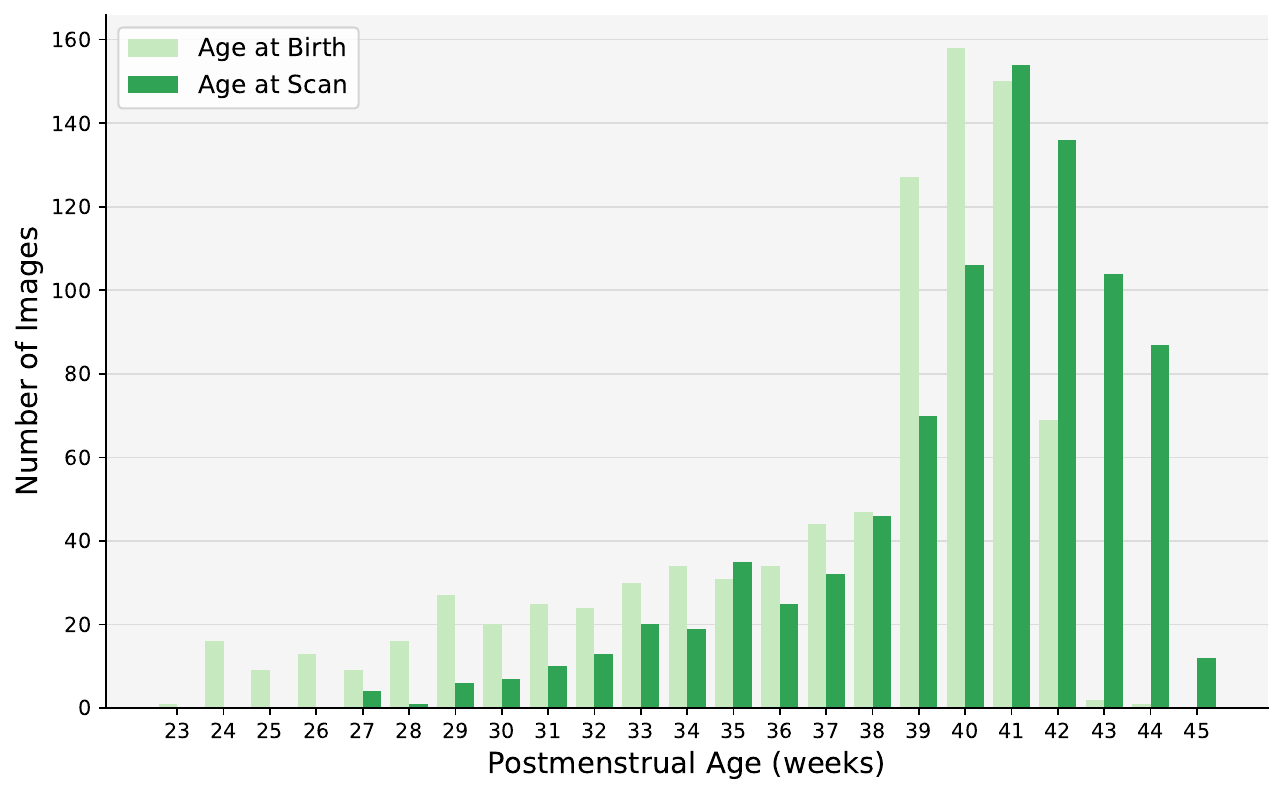}
\caption{The distribution of PMA of neonates at birth and scan.}
\label{fig:age}
\end{figure}

\begin{figure*}[ht]
\centering
\includegraphics[width=1.0\linewidth]{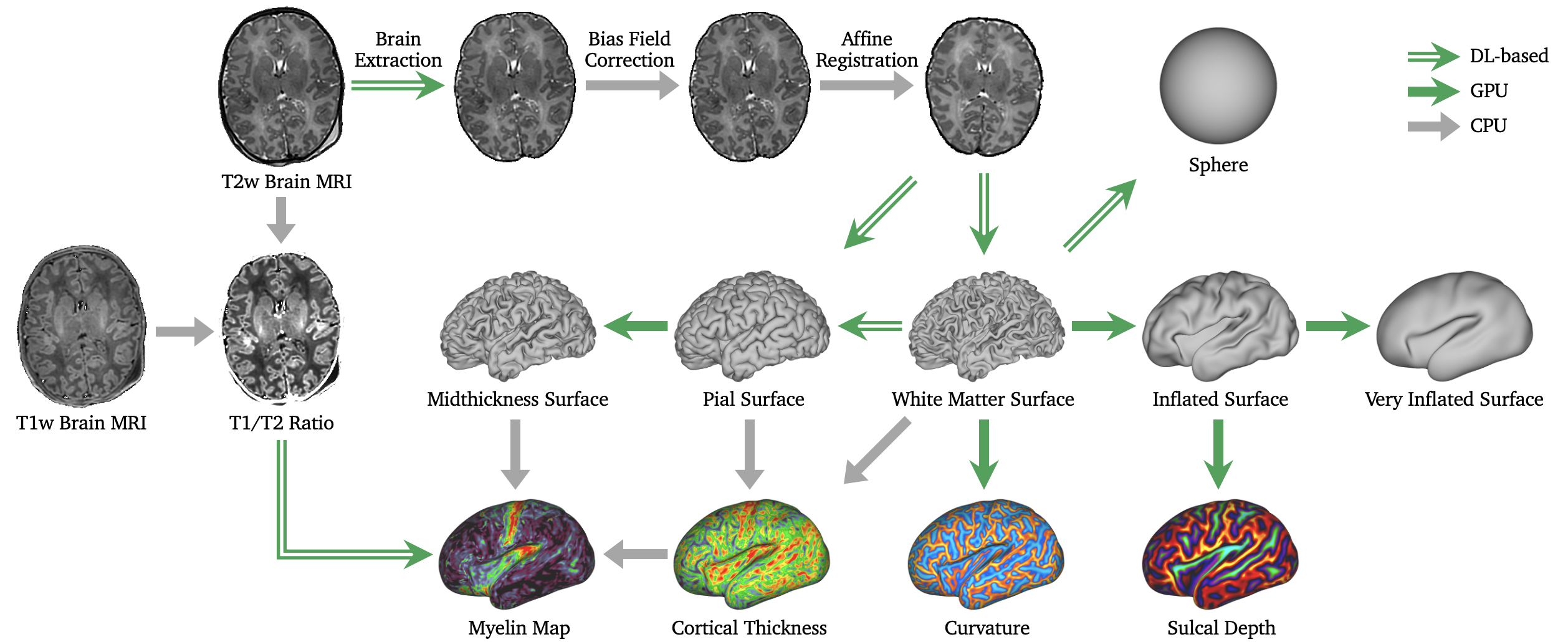}
\caption{The workflow of our DL-based dHCP neonatal brain MRI processing pipeline. The pipeline consists of DL-based, GPU-accelerated and CPU-based processing steps. Given a T2w structural brain MRI, the pipeline learns to extract the brain ROI (Figure \ref{fig:brain_extract}) and applies N4 bias field correction. The processed image is then affinely aligned to a 40-week dHCP atlas. The DL-based pipeline predicts cortical white matter and pial surfaces end-to-end from the aligned image (Figure \ref{fig:surf_recon}). Next, the white matter surface is iteratively inflated and smoothed to produce inflated surfaces. Based on predicted cortical surfaces, the cortical morphological features including cortical thickness, curvature and sulcal depth are computed. The myelin map is estimated if the T1w image is provided. Finally, an unsupervised spherical mapping approach is adopted to project the white matter surface onto a sphere (Figure \ref{fig:sphere_proj}).}
\label{fig:pipeline}
\end{figure*}

\section{Deep Learning-based dHCP Neonatal Pipeline}\label{sec:pipeline}

\subsection{Data Acquisition}\label{subsec:data-aquire}
In this paper, we use the neonatal dataset from the third public dHCP data release\footnote{\url{https://biomedia.github.io/dHCP-release-notes/}} \citep{edwards2022release}. The dHCP neonatal dataset includes 783 neonatal subjects with 887 scans. A small portion of the subjects were scanned more than once for the purpose of longitudinal study. The neonates were scanned at post-menstrual age (PMA) between 27 and 45 weeks and the distribution of their ages at birth and scan is shown in Figure \ref{fig:age}.

For MRI acquisition, the infants were imaged in natural sleep on a 3T Philips scanner with a dedicated neonatal brain imaging system \citep{hughes2017dedicated}. The T2w images are acquired in two stacks of 2D slices in axial and sagittal planes at $0.8\times0.8mm^2$ resolution and $1.6mm$ overlapping slices. The acquisition uses the parameters repetition/echo time TR/TE=12s/156ms and SENSE factor of 2.11 (axial) and 2.60 (sagittal). A motion correction technique \citep{cordero2018motion} is adopted to compensate for head motion during imaging. The acquired images are reconstructed to high-quality 3D volumes at $0.5mm^3$ isotropic resolution \citep{kuklisova2012reconstruction}. The detailed MRI acquisition protocol is described in \cite{edwards2022release}.

\subsection{Pipeline Overview}\label{subsec:overview}

The workflow of our DL-based dHCP neonatal pipeline is illustrated in Figure \ref{fig:pipeline}. Our DL-based pipeline accelerates the original dHCP structural pipeline by integrating several DL-based modules and re-implemented algorithms with GPU acceleration. The pipeline consists of five main steps: preprocessing, cortical surface reconstruction, cortical surface inflation, cortical feature estimation, and spherical projection.

Following the original dHCP structural pipeline \citep{makropoulos2018dhcp}, we use T2w structural brain MRI as the primary inputs. Given a T2w image, the pipeline uses a 3D U-Net to learn a brain mask to extract the brain ROI. Then, N4 bias field correction \citep{tustison2010n4} is applied to normalize the intensity of the T2w image. The processed images are affinely aligned to a 40-week dHCP atlas \citep{schuh2018atlas}. A multiscale deformation network, which learns a sequence of diffeomorphic surface deformations, is introduced to reconstruct cortical white matter and pial surfaces end-to-end from the aligned T2w image. The predicted white matter surface is iteratively inflated and smoothed to generate inflated and very inflated surfaces. Next, the morphological features of the cortex, including cortical thickness, curvature and sulcal depth, are computed from the extracted cortical surfaces. If the T1w MRI is provided, the pipeline estimates the myelin map based on the T1/T2 ratio. Finally, the pipeline employs a Spherical U-Net \citep{zhao2019sphericalunet,zhao2022sphereproj} to project the white matter surface onto a sphere while minimizing the metric distortions.

\begin{figure}[ht]
\centering
\includegraphics[width=1.\linewidth]{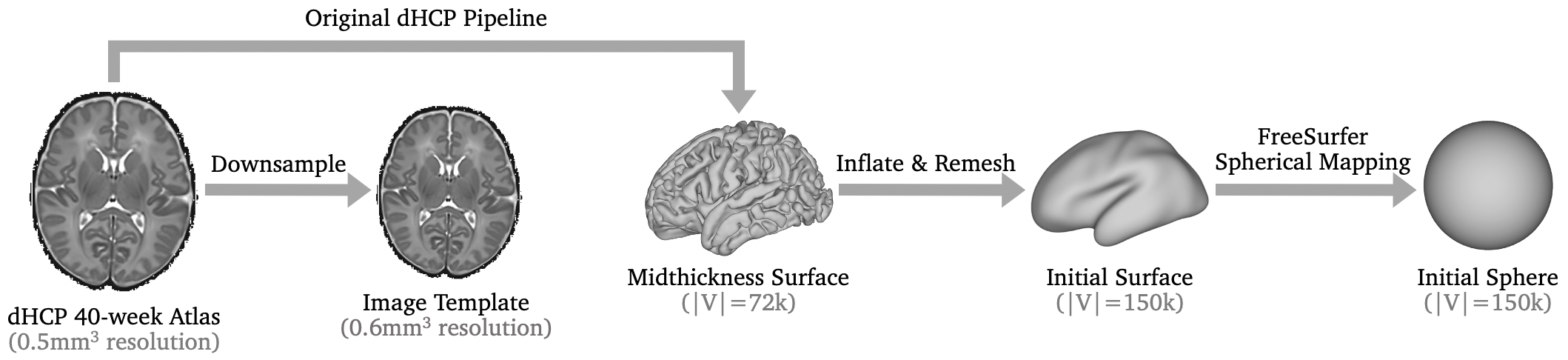}
\caption{The generation of initial image, surface and sphere templates. The 40-week dHCP atlas is downsampled to $0.6mm^3$ resolution as the image template. The midthickness surface for each brain hemisphere, which is extracted from the 40-week dHCP atlas by running the original dHCP pipeline, is inflated and remeshed to create an initial template surface. Then, the initial surface is projected onto an initial sphere through FreeSurfer. $|V|$ is the number of mesh vertices.}
\label{fig:template}
\end{figure}

\subsection{Dataset Preparation}\label{subsec:data-prepare}

To train and evaluate each component of the proposed DL-based pipeline, we use 879 images in the dHCP neonatal dataset, in which 8 samples are excluded during data cleaning. Each data sample includes T2w structural brain MRI as well as pseudo ground truth (GT) segmentations and cortical surfaces generated by the original dHCP structural pipeline \citep{makropoulos2018dhcp}. The original T2w image has the size of $217\times290\times290$. The dataset is randomly split into training/validation/testing sets with the ratio of $60/10/30\%$.

We generate the initial template image, surface and sphere for training and running the pipeline. The generated templates are shared for all subjects. As depicted in Figure~\ref{fig:template}, we create an image template from a 40-week dHCP atlas \citep{schuh2018atlas}. As we aim to develop a fast and memory-efficient pipeline, we downsample the original atlas from $0.5mm^3$ to $0.6mm^3$ isotropic resolution and clip it to the size of $176\times224\times160$ to reduce the GPU memory cost. The downsampled atlas is used as the target image template for affine registration. It should be noted that the downsampling is not necessary if GPU memory allows.

Next, we run the original dHCP pipeline to extract the midthickness surfaces from the dHCP atlas. For each brain hemisphere, the midthickness surface is iteratively smoothed and remeshed to create an inflated initial surface, which is in the same space as the image template. Such an initial surface mesh has 150k vertices and is used as the input template surface for cortical surface reconstruction. This enables the proposed DL-based pipeline to predict cortical surfaces with the same mesh connectivity and fixed number of vertices, while the original dHCP pipeline extracts cortical surfaces with adapted number of vertices, leading to inaccurate estimation of cortical morphological features such as sulcal depth for younger subjects. Finally, we use FreeSurfer \emph{mris\_inflate} and \emph{mris\_sphere} \citep{fischl1999cortical,fischl2012freesurfer} to project the initial surface onto an initial sphere, which has the same mesh connectivity as the initial surface with minimized metric distortions. The initial sphere is used as the input for spherical projection.

\subsection{Preprocessing}\label{subsec:preprocess}

\begin{figure}[ht]
\centering
\includegraphics[width=1.0\linewidth]{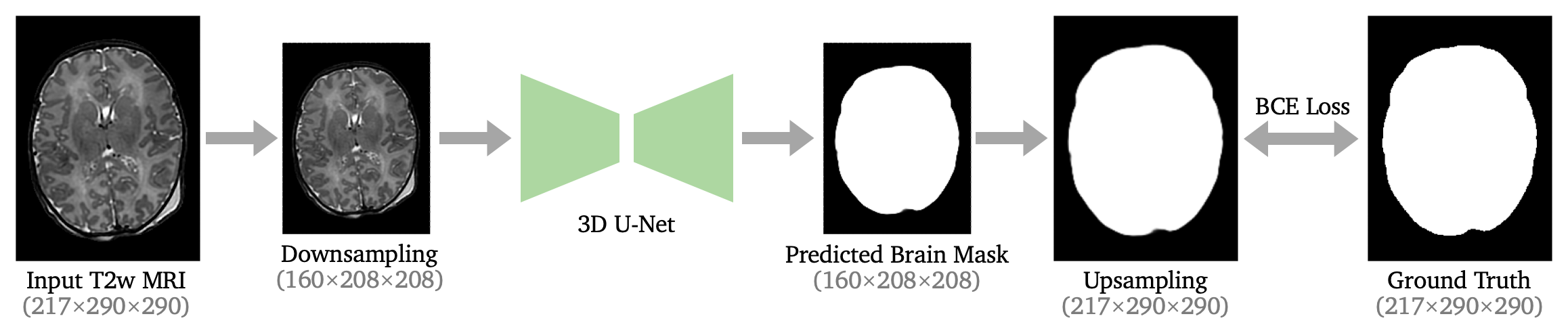}
\caption{The learning-based approach for brain extraction. An original T2w image is downsampled as the input for a 3D U-Net to predict a brain mask. The Binary Cross Entropy (BCE) loss is computed between the upsampled predicted brain mask and the ground truth.}
\label{fig:brain_extract}
\end{figure}

\begin{figure*}[ht]
\centering
\includegraphics[width=1.0\linewidth]{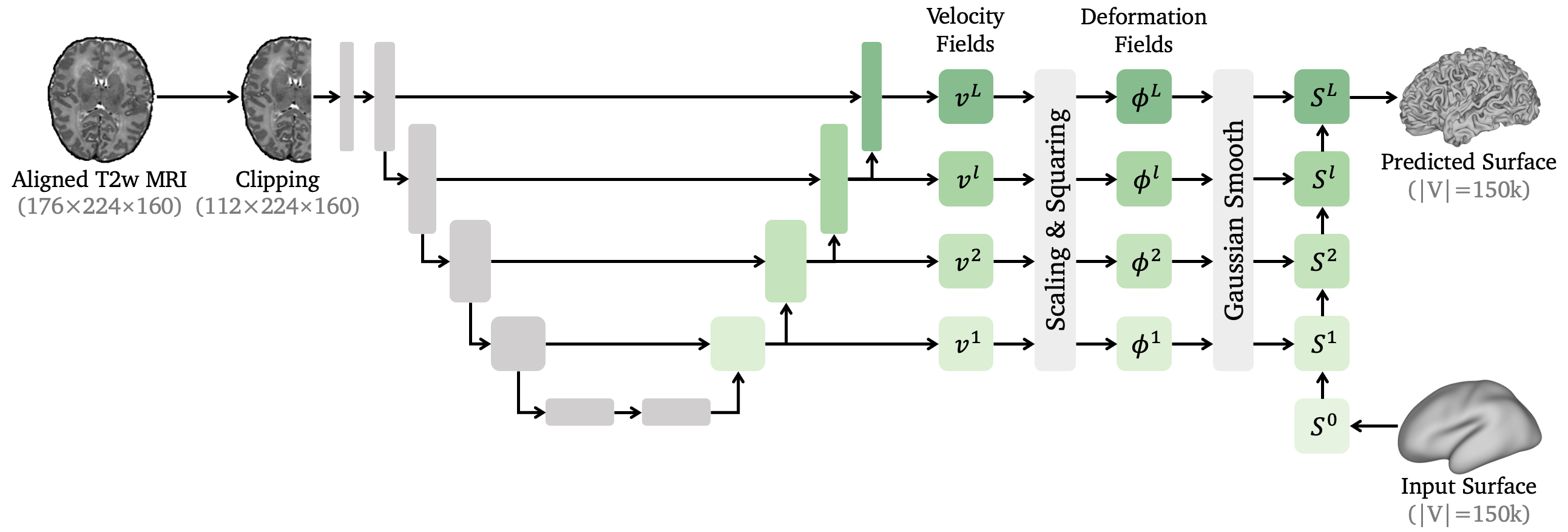}
\caption{The architecture of the multiscale deformation network. An affinely aligned T2w MRI is clipped to the left or right hemisphere as the input. A 3D U-Net is employed to predict multiscale stationary velocity fields (SVF). The SVFs are integrated by scaling and squaring method to multiscale diffeomorphic surface deformation fields, which are smoothed by a Gaussian smoothing layer. The multiscale deformation fields deform an input surface iteratively to the predicted cortical surface, where $|V|$ is the number of mesh vertices.}
\label{fig:surf_recon}
\end{figure*}

\paragraph{Brain Extraction}
The preprocessing steps consist of brain extraction, bias field correction and affine registration. For brain extraction, as shown in Figure \ref{fig:brain_extract}, we train a 3D U-Net \citep{ronneberger2015unet} to predict a brain mask from the input original T2w image. To reduce the GPU memory usage, we downsample the input T2w images to the size of $160\times208\times208$. The predicted brain masks are upsampled to the original size to increase accuracy. We compute the Binary Cross Entropy (BCE) loss between the predicted and GT brain masks. The GT brain masks are generated by FSL BET \citep{smith2002bet}, which is integrated into the original dHCP structural pipeline \citep{makropoulos2018dhcp}. The 3D U-Net model is trained using the Adam optimizer \citep{kingma2014adam} with learning rate $10^{-4}$ for 100 epochs.

\paragraph{Bias Field Correction and Affine Registration}
We use N4 bias field correction \citep{tustison2010n4} to normalize the intensity of the brain-extracted T2w images. Then, we affinely align the bias-corrected image to the dHCP 40-week image template as described in Section~\ref{subsec:data-prepare}. To increase accuracy, we first apply rigid registration to obtain an initial transformation, and then apply the affine registration. The affine registration transforms all MRI into a template space with the size of $176\times224\times160$, which reduces the GPU memory cost for the following process. The proposed pipeline integrates Advanced Normalization Tools in Python (ANTsPy) \citep{avants2009ants}, which provides CPU-based implementation for both N4 algorithm and affine registration.

\subsection{Cortical Surface Reconstruction}\label{subsec:surf-recon}

We propose an end-to-end learning-based approach for cortical surface reconstruction. To tackle the large deformation required by neonatal cortical surfaces, we introduce multiscale deformation networks, which learn a sequence of diffeomorphic transformations to deform an initial surface to white matter and pial surfaces. The architecture of the proposed framework is shown in Figure~\ref{fig:surf_recon}.

\subsubsection{Diffeomorphic Surface Deformation}
Diffeomorphic transformation has been extensively applied to medical image registration \citep{beg2005lddmm,ashburner2007dartel,vercauteren2009demon,balakrishnan2019voxelmorph} and been introduced for learning-based cortical surface reconstruction \citep{lebrat2021corticalflow,santa2022corticalflow,ma2022cortexode,ma2023cotan}. For any domain $\Omega\subset\mathbb{R}^d$, the diffeomorphic deformations can be modeled by the following flow ordinary differential equation (ODE):
\begin{equation}\label{eq:ode}
\frac{\partial}{\partial t}\phi_t=v_t(\phi_t),~\phi_0=id,
\end{equation}
where $v_t:\Omega\times [0,T]\rightarrow\mathbb{R}^d$ is a time-varying velocity field, $\phi_t:\Omega\rightarrow\Omega$ is a deformation field for $t\in[0,T]$, and $\phi_0=id$ is an identity map. In this work, we only consider stationary velocity fields (SVF), \emph{i.e.}, $v_t=v:\Omega\rightarrow\mathbb{R}^d$. If the SVF $v$ is sufficiently smooth, then $\{\phi_t\}_{t\in{[0,T]}}$ represents an one-parameter subgroup of diffeomorphisms generated by $v$ and satisfies the group law $\phi_{s+t}=\phi_{s}\circ\phi_{t}$ \citep{ashburner2007dartel,balakrishnan2019voxelmorph}. Therefore, we can solve the ODE (\ref{eq:ode}), of which the solution is a diffeomorphism $\phi_T$, using the scaling and squaring method \citep{higham2005scaling,arsigny2006scaling,ashburner2007dartel,balakrishnan2019voxelmorph} to recursively compute $\phi_{t}=\phi_{t/2}\circ\phi_{t/2}$. 
For surface deformation, the deformation field $\phi_T$ evolves an initial surface $S_0\subset\mathbb{R}^3$ to a target surface $S_T=\phi_T(S_0)$ while preserving the topology of the surface. Since $\phi_T$ is a differentiable bijection, any two distinct points on the initial surface $S_0$ will not be mapped to the same location, which provides theoretical guarantee for preventing surface self-intersections.

\subsubsection{Multiscale Deformation Network}\label{sec:deform_net}
Since a single SVF has limited representation ability, in this work, we propose multiscale deformation networks to learn a series of diffeomorphic deformation fields $\phi^1,...,\phi^L$ for cortical surface reconstruction. As shown in Figure~\ref{fig:surf_recon}, given an affinely aligned T2w image, we clip it from the original size of $176\times224\times160$ to $112\times224\times160$ for either left or right brain hemisphere. Afterwards, the multiscale deformation network predicts cortical surfaces end-to-end from the clipped brain MRI. Since the proposed neural network model only processes one brain hemisphere at a time, the clipped brain MRI not only provides necessary information for cortical surface reconstruction, but also effectively reduces both GPU memory costs and inference time. Note that the brain segmentations are not required as the input, which would cause subsequent corruptions on the surfaces in cases where segmentations are inaccurate.

We use a 3D U-Net to extract multiscale feature maps from the input clipped T2w image. Then, a sequence of multiscale volumetric SVFs $v^1,...,v^L$ are predicted from the extracted feature maps. Each SVF $v^l$ downsamples the original brain MRI volume by a factor of $2^{L-l-1}$ for $l=1,...,L-1$. The last SVF $v^L$ has the same size as the input volume to capture fine details. Previous work \citep{lebrat2021corticalflow,ma2022cortexode} adopted the forward Euler method to solve the ODE (\ref{eq:ode}), which is fast but may introduce large numerical errors. Instead, we use the scaling and squaring method \citep{higham2005scaling,arsigny2006scaling} for integration, which is more accurate and widely used for medical image registration tasks \citep{ashburner2007dartel,balakrishnan2019voxelmorph}. For each scale $l=1,...,L$, we set the initial transformation as $\phi^l_{T/2^K}=id + \frac{T}{2^{K}} v^l$, and then the deformation field can be updated recurrently by $\phi^l_{T/2^{K-1}}=\phi^l_{T/2^K}\circ\phi^l_{T/2^K}$, where $K$ is the steps of the scaling and squaring. This is equivalent to $2^K$ steps of forward Euler.

\begin{figure}[ht]
\centering
\includegraphics[width=1.0\linewidth]{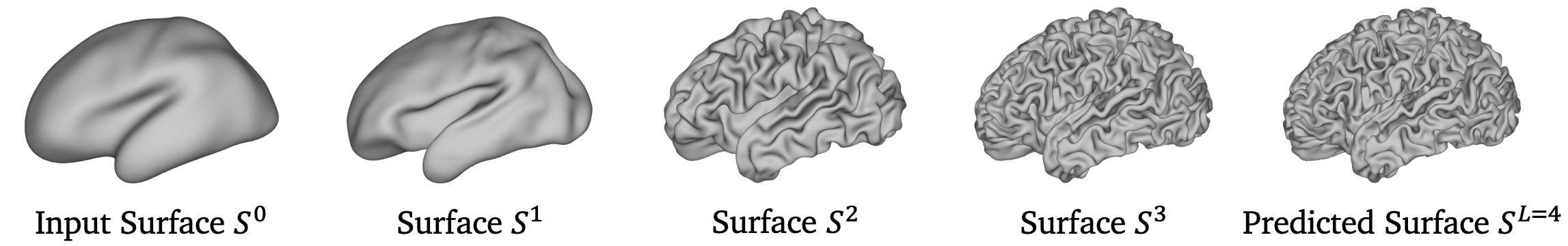}
\caption{The deformed surfaces $S^l$ generated by the multiscale deformation network for $l=0,1,...,L$ with $L=4$, where $S^0$ is the input surface and $S^L$ is the final predicted surface.}
\label{fig:surf_deform}
\end{figure}

By integration, the network predicts multi-resolution diffeomorphic deformation fields $\phi^l$, each of which has the same size as the corresponding SVF $v^l$. To improve the smoothness of the deformation fields, we further employ Gaussian smoothing with standard deviation $\sigma=1.0$. Finally, we apply the deformation fields iteratively to an input surface $S^0\subset\mathbb{R}^3$. The displacements of the surface mesh vertices are sampled by linear interpolation. The surfaces are updated by $S^{l}=\phi^{l}(S^{l-1})$ for $l=1,...,L$. We set the number of scales to $L=4$, which allows the multiscale deformation network to model coarse-to-fine surface deformations. The deformation fields $\phi^1$ and $\phi^2$ have been downsampled by a factor of 4 and 2 respectively. As shown in Figure~\ref{fig:surf_deform}, $\phi^1$ and $\phi^2$ contain global information and deform the input surface $S^0$ into coarse and smooth intermediate surfaces. $\phi^3$ and $\phi^4$, which have the same size as the input brain MRI, capture detailed cortical folds and provide further refinement for the final predicted white matter or pial surface $S^L$.

For white matter surface extraction, we use an initial template surface described in Figure~\ref{fig:template} as the input surface for all subjects. The predicted white matter surfaces are further refined by Taubin smoothing \citep{taubin1995smooth} to improve the mesh quality, \emph{i.e.}, the smoothness and regularity of the triangular mesh. For pial surface reconstruction, we use the predicted white matter surface as the input surface. Since the initial surface has genus-0 topology and the diffeomorphic deformations are topology-preserving, all reconstructed cortical surfaces have the same topology as a 2-sphere.

\subsubsection{Model Training}
We train four multiscale deformation networks for white matter and pial surface reconstruction for both brain hemispheres. We consider the following loss function:
\begin{equation}\label{eq:surf-loss}
\mathcal{L}=\mathcal{L}_{cd} + \lambda_{edge}\mathcal{L}_{edge} + \lambda_{nc}\mathcal{L}_{nc},
\end{equation}
where $\mathcal{L}_{cd}$ is the bidirectional Chamfer distance \citep{fan2017chamfer,wang2018pixel2mesh,wickramasinghe2020voxel2mesh}, $\mathcal{L}_{edge}$ is the edge length loss \citep{wang2018pixel2mesh,wickramasinghe2020voxel2mesh}, $\mathcal{L}_{nc}$ is the normal consistency loss \citep{wang2018pixel2mesh,wickramasinghe2020voxel2mesh,ma2023cotan}, $\lambda_{edge}$ and $\lambda_{nc}$ are weights for regularization terms. The Chamfer distance $\mathcal{L}_{cd}$ is defined as the bidirectional distance between two point sets $X,Y\subset\mathbb{R}^3$ \citep{fan2017chamfer}:
\begin{equation}\label{eq:chamfer-loss}
\mathcal{L}_{cd}(X,Y) = \frac{1}{2|X|}\sum_{x\in X}\min_{y\in Y}\|x-y\|^2 + \frac{1}{2|Y|}\sum_{y\in Y}\min_{x\in X}\|x-y\|^2.
\end{equation}
Specifically, $X$ and $Y$ are the sets of mesh vertices of the predicted and pseudo GT cortical surfaces respectively. The pseudo GT cortical surfaces are generated using the original dHCP pipeline \citep{makropoulos2018dhcp}. All pseudo GT surfaces are remeshed to have 150k vertices to reduce the errors introduced by point matching during the computation of the Chamfer distance. We adopt PyTorch3D \citep{ravi2020pytorch3d} which provides fast computation and gradient backpropagation for the Chamfer distance.

Edge length and normal consistency constraints are introduced to encourage the smoothness and improve the mesh quality of the predicted cortical surfaces. For a surface mesh $M=(V,E,F)$, where $V,E,F$ are the sets for vertices, edges and faces respectively, the edge length loss is defined as:
\begin{equation}\label{eq:chamfer-loss}
\mathcal{L}_{edge}(M)=\frac{1}{|E|}\sum_{(i,j)\in E}\|v_i-v_j\|^2,
\end{equation}
where $v_i,v_j$ are two vertices connected by the edge $e_{ij}=(i,j)$. The normal consistency loss is defined as:
\begin{equation}\label{eq:chamfer-loss}
\mathcal{L}_{nc}(M)=1-\frac{1}{|E|}\sum_{f_i\cap f_j\in E}n_i\cdot n_j.
\end{equation}
Here $n_i$ is the unit normal vector of the face $f_i\in F$, and $e=f_i\cap f_j$ is the edge where the adjacent faces $f_i$ and $f_j$ intersect. The normal consistency loss penalizes the dot product of the normal vectors between any two adjacent faces. The value of the normal consistency loss is determined by the PMA of neonates and the curvature of cortical surfaces. For older neonates with highly folded and curved cortical surfaces, there is a large difference between the normal orientations of two adjacent faces. Consequently, the normal consistency loss reaches a high value to enforce surface smoothness. For younger subjects with immature cortex, the loss is relatively lower as the cortical surfaces are already sufficiently smooth.

For the architecture of the multiscale deformation network, we set $L=4$ as the number of scales as discussed in Section \ref{sec:deform_net}. A smaller $L$ would offer insufficient global information, while a larger $L$ is more suitable for brain MRI at higher image resolution. The numbers of channels for hidden layers are set to $\{16,32,64,128,128\}$ for white matter surface reconstruction and $\{16,32,32,32,32\}$ for pial surface to avoid overfitting. We use the same hyperparameters for both white matter and pial surface reconstruction. The regularization weights are selected by grid search, and set to $\lambda_{edge}=0.3$ and $\lambda_{nc}=3.0$ to achieve satisfactory mesh quality. The number of scaling and squaring steps is set to $K=7$ following \cite{balakrishnan2019voxelmorph}. Empirically \citep{ma2022cortexode,ma2023cotan}, we use the Adam optimizer with learning rate $10^{-4}$ to train the multiscale deformation networks for 200 epochs. After training, we select the best models with minimum validation error, \emph{i.e.}, minimum Chamfer distances between predicted and pseudo GT cortical surfaces.

\begin{figure*}[ht]
\centering
\includegraphics[width=1.0\linewidth]{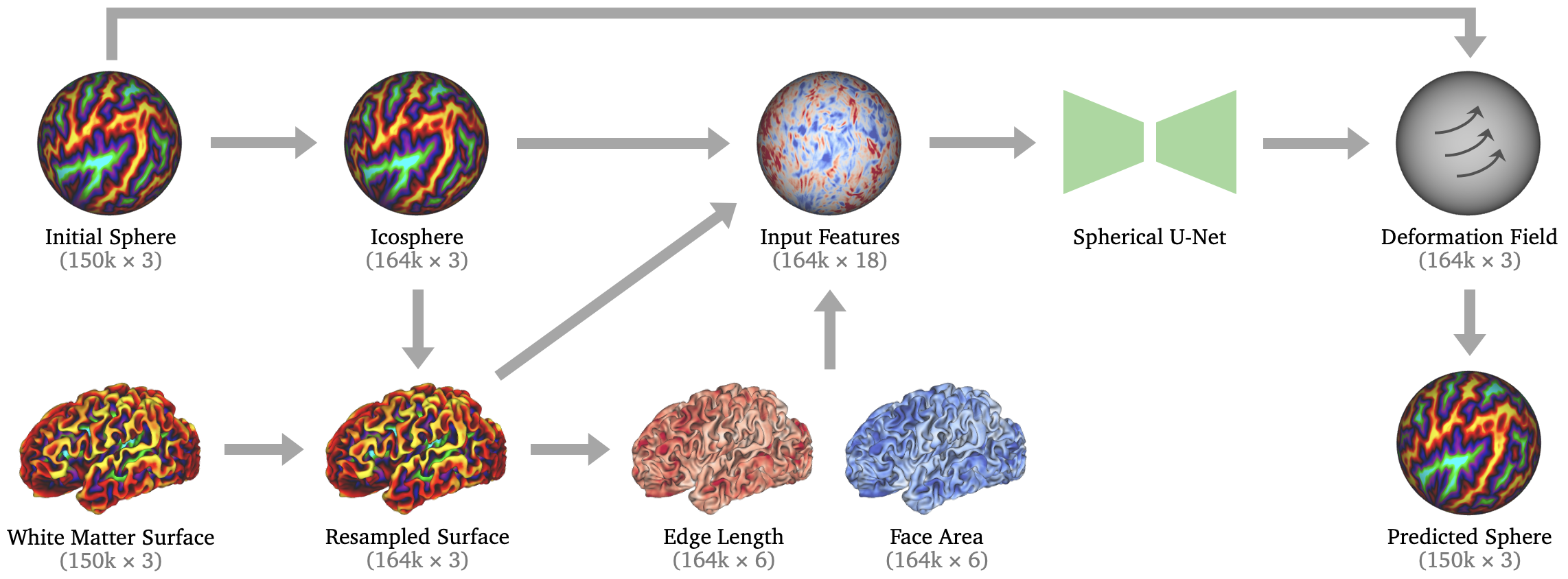}
\caption{Our learning-based spherical projection framework. Given an input white matter surface and an initial sphere, we resample the white matter surface mesh such that it has the same connectivity as a standard icosphere with 164k vertices. Then, we compute the metrics including edge length and face area of the resampled surface mesh, of which each vertex contains the metrics of its adjacent edges or faces. The computed metrics are concatenated with the coordinates of the resampled surface and the icosphere as input features. A Spherical U-Net learns from the input features to predict a diffeomorphic deformation field, which deforms the initial sphere to a predicted sphere. An unsupervised loss function is defined as the metric distortion between the predicted sphere and the original white matter surface.}
\label{fig:sphere_proj}
\end{figure*}

\subsection{Cortical Surface Inflation}\label{subsec:surf-inflate}

Following the HCP \citep{glasser2013hcp} and original dHCP pipelines \citep{makropoulos2018dhcp}, our dHCP DL pipeline incorporates two types of cortical surface inflation. Firstly, the original pipeline employs Connectome Workbench \citep{glasser2013hcp} to iteratively inflate and smooth the midthickness surface to generate inflated and very inflated surfaces. These surfaces are primarily used for visualization. Our pipeline re-implements these surface inflation approaches using PyTorch to facilitate GPU acceleration. 
Secondly, the original dHCP pipeline adopts FreeSurfer \textit{mris\_inflate} command \citep{fischl1999cortical,fischl2012freesurfer}, which is reproduced in the Medical Image Registration ToolKit (MIRTK) \citep{schuh2017deformable,makropoulos2018dhcp}, to inflate the white matter surface while minimizing the metric distortion between white matter and inflated surfaces. Our DL-based pipeline also accelerates such an inflation process by PyTorch with parallel computation on a GPU. 

During the inflation, the cortical sulcal depth, or average convexity of each mesh vertex is computed by integrating the displacement of the vertex along its normal direction. 
{As all cortical surfaces have fixed number of vertices, our DL-based dHCP pipeline provides age-consistent estimation of sulcal depth, while original dHCP pipeline extracts cortical surfaces with less number of vertices for younger infants, resulting in over-inflation of surfaces and overestimation of sulcal depth.}

\subsection{Cortical Feature Estimation}\label{subsec:feat-est}
Based on the extracted cortical surfaces, our DL-based dHCP pipeline provides the estimation of cortical morphological features such as cortical thickness, curvature and sulcal depth, as well as functional features such as a myelin map. The estimation of cortical features is consistent with previous pipelines \citep{glasser2013hcp,makropoulos2018dhcp}. The cortical thickness is computed by averaging the bidirectional distances between white matter and pial surfaces. This is implemented with the SciPy package \citep{virtanen2020scipy} in Python. Furthermore, the mean curvature of the white matter surface is computed via PyTorch-based re-implementation \citep{maillot1993curv,glasser2013hcp}. The sulcal depth is measured during cortical surface inflation as described in Section \ref{subsec:surf-inflate}.

For the myelin map estimation, if the T1 image is provided and aligned to the T2 image, we divide the T1 image by the original T2 image before intensity normalization to compute the T1/T2 ratio. Then, the myelin map is estimated by projecting the T1/T2 ratio onto the midthickness surface using volume-to-surface mapping \citep{glasser2011myelin,glasser2013hcp,makropoulos2018dhcp}, which is implemented with the Connectome Workbench. For each vertex of the midthickness surface mesh, we find its corresponding location in the T1/T2 ratio volume, and compute the Gaussian average of its neighboring voxels as the myelin map value, where the size of the neighbor depends on the cortical thickness. Only voxels within the cortical ribbon ROI are considered. However, it is computationally intensive to examine if a voxel or a point is located between the inner and outer cortical surfaces, in particular for high resolution meshes with 150k vertices. To accelerate the volume-to-surface mapping, we train a 3D U-Net to learn the binary cortical ribbon mask, \emph{i.e.}, the cortical GM segmentation. The learning procedure is similar to brain extraction as shown in Figure \ref{fig:brain_extract}. The pseudo GT segmentation is generated by the Draw-EM approach integrated in the original dHCP pipeline.

\subsection{Spherical Projection}\label{subsec:sphere-proj}

The purpose of spherical projection or spherical mapping is to project the white matter surface onto a sphere while minimizing metric distortions. Motivated by \cite{zhao2022sphereproj}, we propose an unsupervised learning framework for spherical projection and integrate it into our DL-based dHCP pipeline. As illustrated in Figure \ref{fig:sphere_proj}, the inputs of our spherical projection framework include a white matter surface and a fixed initial sphere for all subjects. Note that since all white matter surfaces are deformed from the same initial template surface, the white matter surfaces have the same mesh connectivity as the initial sphere, which is generated from the initial surface by FreeSurfer (see Figure \ref{fig:template}). Therefore, instead of inflating the white matter surface and projecting the inflated surface to create a sphere for each subject \citep{fischl1999cortical,elad2005mds,zhao2022sphereproj}, our approach learns to deform the initial sphere directly while minimizing metric distortions between the deformed sphere and the white matter surface. In this work, we primarily consider edge and area distortions.

We resample the input white matter surface mesh such that it has the same connectivity as a standard icosphere with 163,842 vertices. This is achieved by barycentric interpolation \citep{robinson2014msm,robinson2018msm} between the initial sphere and the icosphere, of which the barycentric coordinates are pre-computed to accelerate the resampling. Then, we compute the metrics of the resampled white matter surface including edge length and face area. Each vertex contains the metrics of its adjacent edges and faces. We concatenate the coordinates of the resampled white matter surface and the icosphere as well as computed metrics as input features of a Spherical U-Net \citep{zhao2019sphericalunet}, which has been widely used in cortical surface analysis tasks such as cortical surface registration \citep{zhao2021s3reg}, parcellation \citep{zhao2019parcellation} and development prediction \citep{zhao2021develop}. We train the Spherical U-Net to learn a diffeomorphic deformation field in the spherical domain by integrating a SVF using scaling and squaring \citep{higham2005scaling,arsigny2006scaling}. Rather than adding regularization terms to the loss function \citep{zhao2022sphereproj}, we apply Laplacian mesh smoothing to the deformation field to encourage smoothness. The predicted sphere is derived by deforming the initial sphere, where the displacement of each vertex is determined by sampling from the deformation field using barycentric interpolation.

\cite{zhao2022sphereproj} employed multi-resolution Spherical U-Nets and computed the losses between the deformed icosphere and the resampled white matter surface with 164k vertices. In contrast, our approach minimizes the metric distortions end-to-end between the predicted sphere and the original input white matter surface. Hence, a single-scale Spherical U-Net is sufficient to achieve accurate results. Inspired by \cite{zhao2022sphereproj}, we train the Spherical U-Net by minimizing an unsupervised metric distortion loss, which is defined as the minimum root mean square deviation (RMSD) between the metrics of predicted sphere and the original white matter surface:
\begin{equation}\label{eq:distortion}
\mathcal{L}(X_{sphere}, X_{surf})=\min_{\beta\in\mathbb{R}_+}\sqrt{\frac{1}{N}\sum_{i=1}^N\left(\beta X^i_{sphere}-X^i_{surf}\right)^2},
\end{equation}
where $X_{sphere},X_{surf}\in\mathbb{R}^N$ are the metrics and $\beta$ is a scaling coefficient. The loss (\ref{eq:distortion}) achieves minimum when
\begin{equation}\label{eq:beta}
\beta=\sum_{i=1}^N \left(X^i_{sphere}X^i_{surf}\right)/\sum_{i=1}^N \left(X^i_{sphere}\right)^2.
\end{equation}
The metrics in Eq.~(\ref{eq:distortion}) can be replaced by the mesh edge length or face area. The overall loss function is defined as $\mathcal{L}=\lambda_{e}\mathcal{L}_{e} + \lambda_{a}\mathcal{L}_{a}$, where $\mathcal{L}_{e},\mathcal{L}_{a}$ are the edge and area distortions, and $\lambda_{e},\lambda_{a}$ are weights for distortions. For model training, we use the Adam optimizer with learning rate $10^{-4}$ to train for 200 epochs. The weights of losses are set to $\lambda_e=1.0$ and $\lambda_a=0.5$. After training, we select the model with lowest metric distortions on the validation set.

\begin{table*}[h]
\centering\small
\begin{tabular}{lcc}
\toprule
& Original dHCP Pipeline & DL-based dHCP Pipeline (Ours)\\ 
\midrule
Brain Extraction & FSL BET \citep{smith2002bet} & Supervised Learning + U-Net\\
Tissue Segmentation & Draw-EM \citep{makropoulos2014drawem} & --- \\
Cortical Surface Reconstruction & Deformable Model \citep{schuh2017deformable} & Multiscale Deformation Networks\\
Spherical Projection & Spherical MDS \citep{elad2005mds} & Unsupervised Learning + Spherical U-Net  \\
\bottomrule
\end{tabular}
\caption{Comparisons of the approaches for brain extraction, tissue segmentation, cortical surface reconstruction and spherical projection between the original dHCP pipeline \citep{makropoulos2018dhcp} and our DL-based dHCP pipeline.}
\label{tab:compare_approach}
\end{table*}

\begin{table}[h]
\centering\small
\begin{tabular}{lccc}
\toprule
Processing Steps & Implementation & GPU & CPU only  \\
\midrule
Brain Extraction & PyTorch &  0.25s & 1.63s  \\
Bias Field Correction & ANTsPy & 2.62s & 2.64s \\
Affine Registration & ANTsPy & 2.76s & 2.77s \\
Surface Reconstruction & PyTorch3D & 2.37s & 37.01s \\
Surface Inflation & PyTorch & 4.99s  & 21.28s \\
Spherical Projection & PyTorch & 0.68s & 1.35s \\
% \cline{3-4}
Cortical Thickness & SciPy & \multirow{4}{*}{8.73s} & \multirow{4}{*}{102.15s}\\
Mean Curvature & PyTorch & & \\
Sulcal Depth & PyTorch & & \\
Myelin Map & WB \& PyTorch & & \\
\midrule
Total Runtime & & 23.98s & 170.29s \\
\bottomrule
\end{tabular}
\caption{The implementation and runtime of the proposed dHCP DL pipeline. Both GPU-accelerated and CPU-only runtime (without GPU acceleration) are reported. In addition to listed processing steps, the total runtime also involves extra processing time such as file I/O operations.}
\label{tab:compare_perform}
\end{table}

\begin{table}[h]
\centering\small
\begin{tabular}{lccc}
\toprule
Processing Steps & FastSurfer & iBEAT V2.0 & Ours\\
\midrule
Preprocessing &  & \cmark  & \cmark \\
Brain Extraction &  & {\color{green}\cmark}  & {\color{green}\cmark} \\
Brain Segmentation & {\color{green}\cmark} & {\color{green}\cmark} &  \\
Surface Reconstruction & \cmark & \cmark & {\color{green}\cmark} \\
Surface Parcellation & \cmark & \cmark &  \\
Spherical Projection & \cmark &  & {\color{green}\cmark} \\
Feature Estimation & \cmark & \cmark & \cmark \\
\midrule
Approx Runtime & 1h & 4h & \textbf{24}s \\
\multirow{2}{*}{CPU Specs (Intel)} & Xeon & Core
 & Core \\
& Gold 6154 & i7-8700k &  i7-11700K \\
GPU Specs (NVIDIA) & Titan Xp & 1080 Ti & 3080 \\
\bottomrule
\end{tabular}
\caption{Comparisons of functionality and runtime among DL-based neuroimage processing pipelines: FastSurfer, iBEAT V2.0, and our dHCP DL pipeline. The green check marks indicate DL-based processing steps. We provide the approximated runtime as reported in original papers \citep{henschel2020fastsurfer,wang2023ibeat}. The CPU and GPU specifications used in the original papers are reported as well for a fair comparison.}
\label{tab:compare_learning}
\end{table}

\section{Comparison to Existing Neuroimage Pipelines}\label{sec:compare}

\subsection{Comparison to Original dHCP Pipeline}\label{sec:compare_dhcp}
We compare our DL-based dHCP neonatal pipeline to the original dHCP pipeline \citep{makropoulos2018dhcp} with respect to the approach, implementation and runtime. The comparison of surface and sphere quality are provided in the Section \ref{sec:qc}.

\subsubsection{Approach and Implementation}\label{subsubsec:compare_impl}
We compare the different approaches of the main processing steps between the original and our DL-based dHCP pipelines in Table \ref{tab:compare_approach}. Our DL-based pipeline integrates novel learning-based approaches for the crucial processing steps, and predicts cortical surfaces end-to-end without the need of tissue segmentation.

The original dHCP pipeline is implemented with MIRTK \citep{makropoulos2018dhcp}, Connectome Workbench \citep{glasser2013hcp}, and FSL \citep{jenkinson2012fsl}. Since the installation and configuration of the MIRTK package are complicated, it is difficult to reproduce or modify the original dHCP pipeline. For our DL-based dHCP pipeline, the implementation of each processing step is described in Table~\ref{tab:compare_perform}. Our DL-based dHCP pipeline is primarily implemented in Python and PyTorch \citep{paszke2019pytorch}, which simplify deployment, enhance reproducibility, and allow for more straightforward modifications and adaptations. The PyTorch library provides seamless support for parallel computation and acceleration on GPUs. The PyTorch3D package \citep{ravi2020pytorch3d} is required for the training of cortical surface reconstruction and provides utilities for mesh-based processing. ANTsPy \citep{avants2009ants} is used for N4 bias correction and affine registration. The Connectome Workbench (WB) is used for visualization and myelin map estimation. 
Our DL-based pipeline is GPU memory-efficient, which only uses 12GB for training and 7.6GB GPU memory for inference. Therefore, it is simple to adapt the pipeline to a new dataset by fine-tuning the neural network models.

\subsubsection{Runtime}\label{subsubsec:compare_time}
To measure the runtime, we run both original and DL-based dHCP pipeline to process the brain MRI of 10 dHCP neonatal subjects, which are randomly selected from the test set. The runtime is measured on an 8-core Intel Core i7-11700K CPUs and a NVIDIA GeForce RTX 3080 GPU with 12GB memory. The original dHCP pipeline \citep{makropoulos2018dhcp} runs exclusively on CPUs and requires 6h38min to process a single subject. The volume-based and surface-based processing requires 2h9min and 4h29min respectively. For our DL-based pipeline, we report the runtime with and without GPU acceleration in Table~\ref{tab:compare_perform}. With GPU acceleration, it only takes 24 seconds to run the entire DL-based pipeline, which is 995$\times$ faster than the original dHCP pipeline. The cortical surface reconstruction and spherical projection only need 3 seconds in total owing to the powerful end-to-end deep learning approaches. Even without GPU acceleration, our DL-based pipeline only requires 170 seconds to complete on the CPU, which is still 140$\times$ faster compared to the original dHCP pipeline.

\begin{figure}[ht]
\centering
\includegraphics[width=0.9\linewidth]{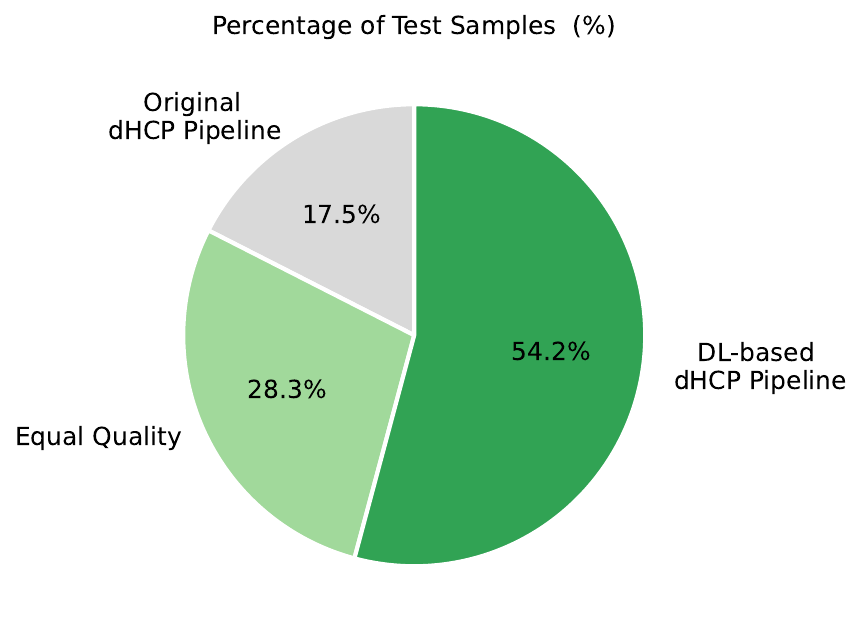}
\includegraphics[width=0.9\linewidth]{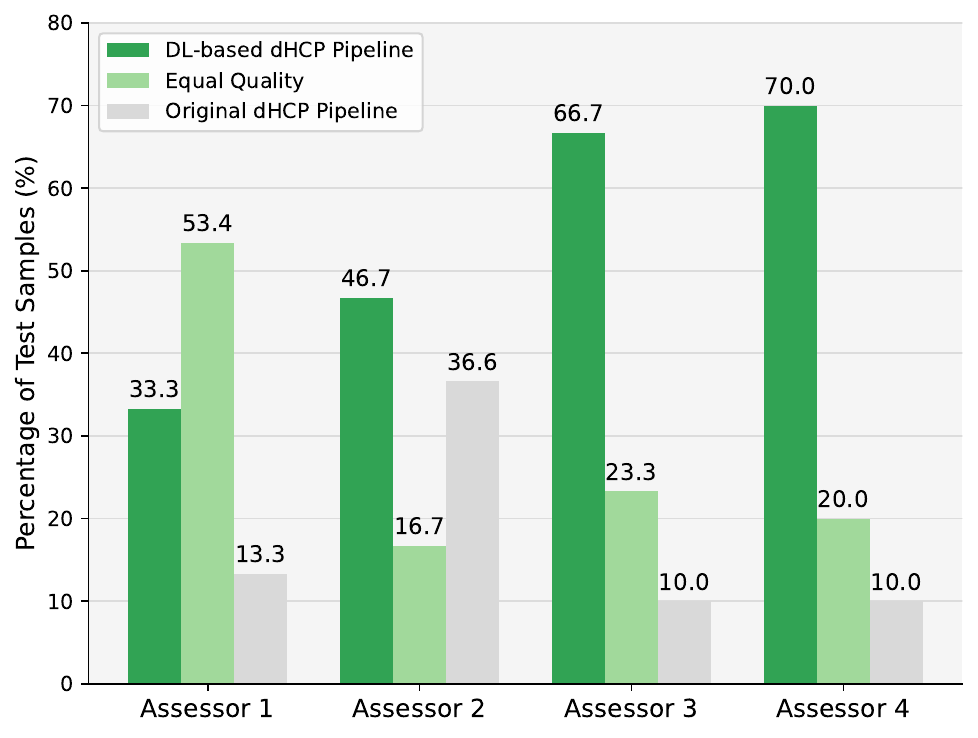}
\caption{The results of manual cortical surface quality control (QC). The assessors visually examine and compare the anatomical correctness of cortical surfaces between the original and our DL-based dHCP pipelines for each test brain MRI. Top: the average surface QC results. For 82.5\% of all test samples, our DL-based dHCP pipeline achieves superior (54.2\%) or equal (28.3\%) surface quality compared to the original dHCP pipeline. Bottom: the individual surface QC results from all assessors.}
\label{fig:qc}
\end{figure}

\begin{figure*}[ht]
\centering
\includegraphics[width=0.95\linewidth]{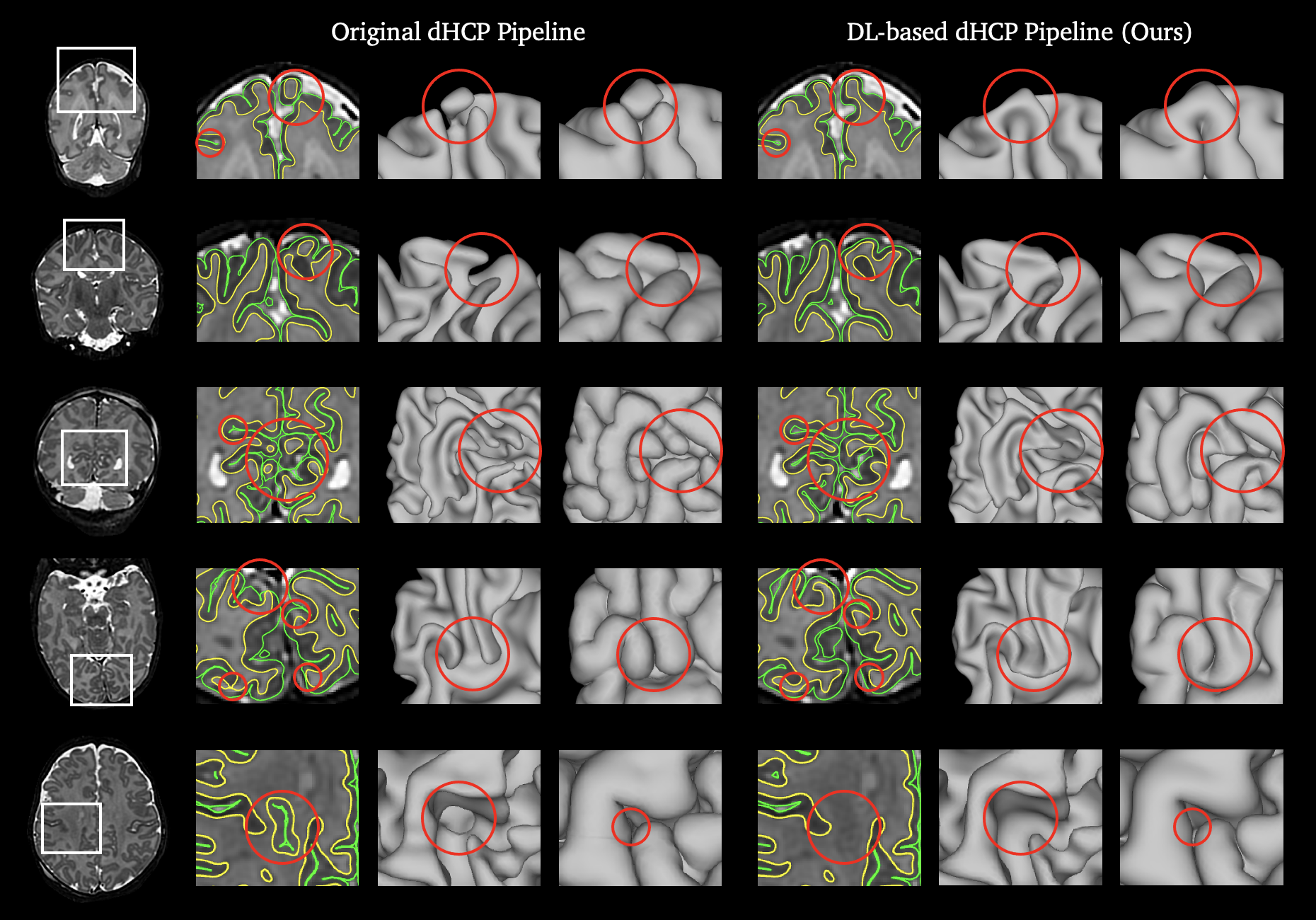}
\caption{The qualitative comparison between original and our DL-based dHCP pipeline on 5 different test samples of the dHCP dataset. The left columns visualize the corruptions in cortical surfaces produced by the original dHCP pipeline, while the right columns show that our DL-based dHCP pipeline effectively reduces these errors. The yellow boundary is the white matter surface, and the green boundary is the pial surface. Row 1-2: corruptions on the top of the cortical surfaces. Row 2-4: corruptions on the back of the cortical surfaces. Row 5: the "cave/hole" introduced by the errors of WM segmentations.}
\label{fig:surf_quality}
\end{figure*}

\begin{figure}[h]
\centering
\includegraphics[width=1.0\linewidth]{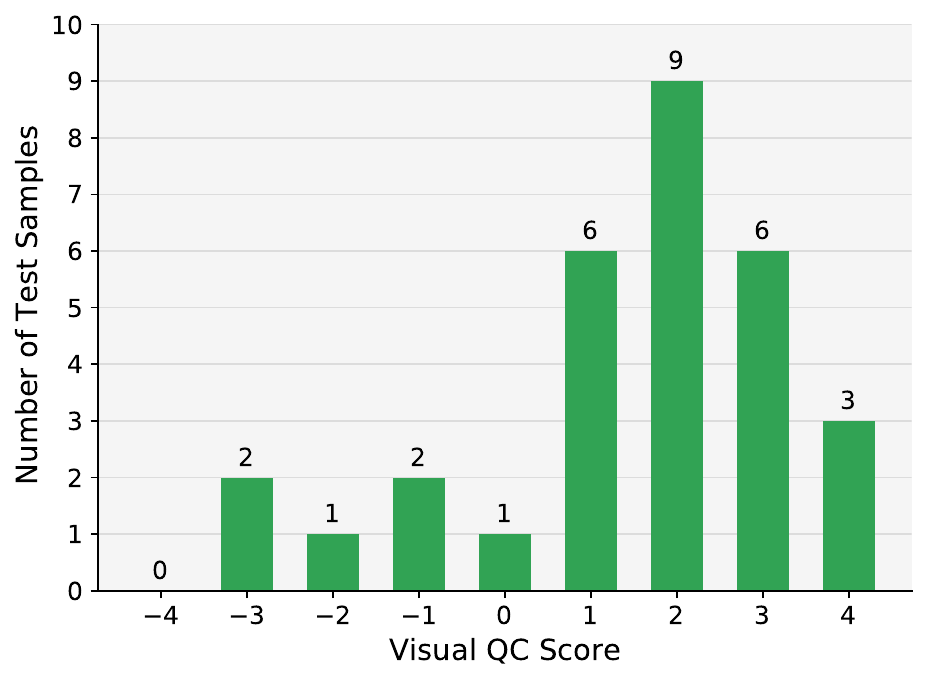}
\caption{The number of test samples grouped by the visual QC scores.}
\label{fig:qc_score}
\end{figure}

\begin{figure*}[ht]
\centering
\includegraphics[width=1.0\linewidth]{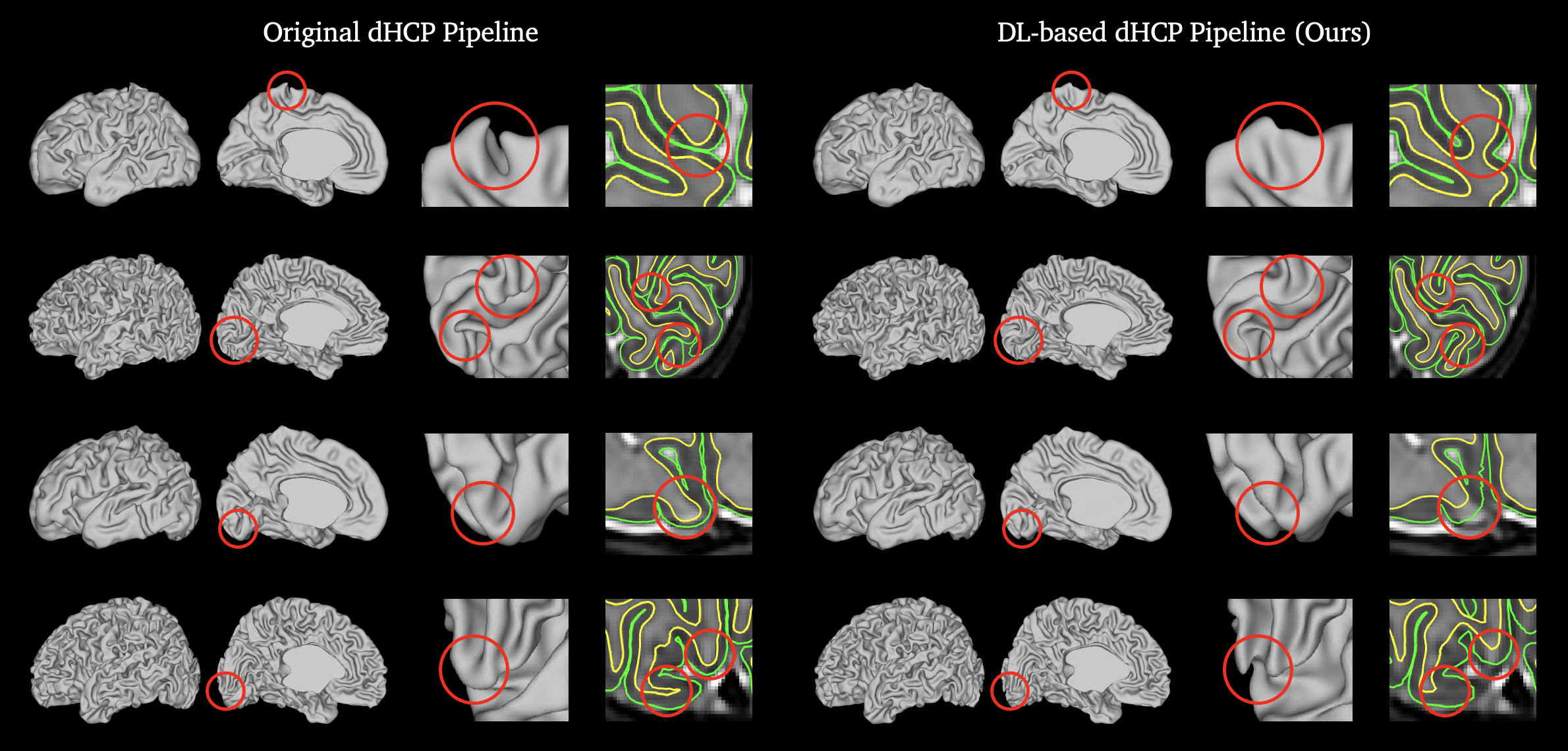}
\caption{The visualization of succeeded and failed cases compared between original dHCP pipeline and our DL-based pipeline. Row 1-2: cortical surfaces of two test samples when 4 assessors agree that our DL-based pipeline produces better surface quality. Row 3-4: cortical surfaces of two test samples when 4 assessors agree that the surface quality of the original dHCP pipeline is no worse than ours.}
\label{fig:fail_case}
\end{figure*}

\begin{figure}[ht]
\centering
\includegraphics[width=1.0\linewidth]{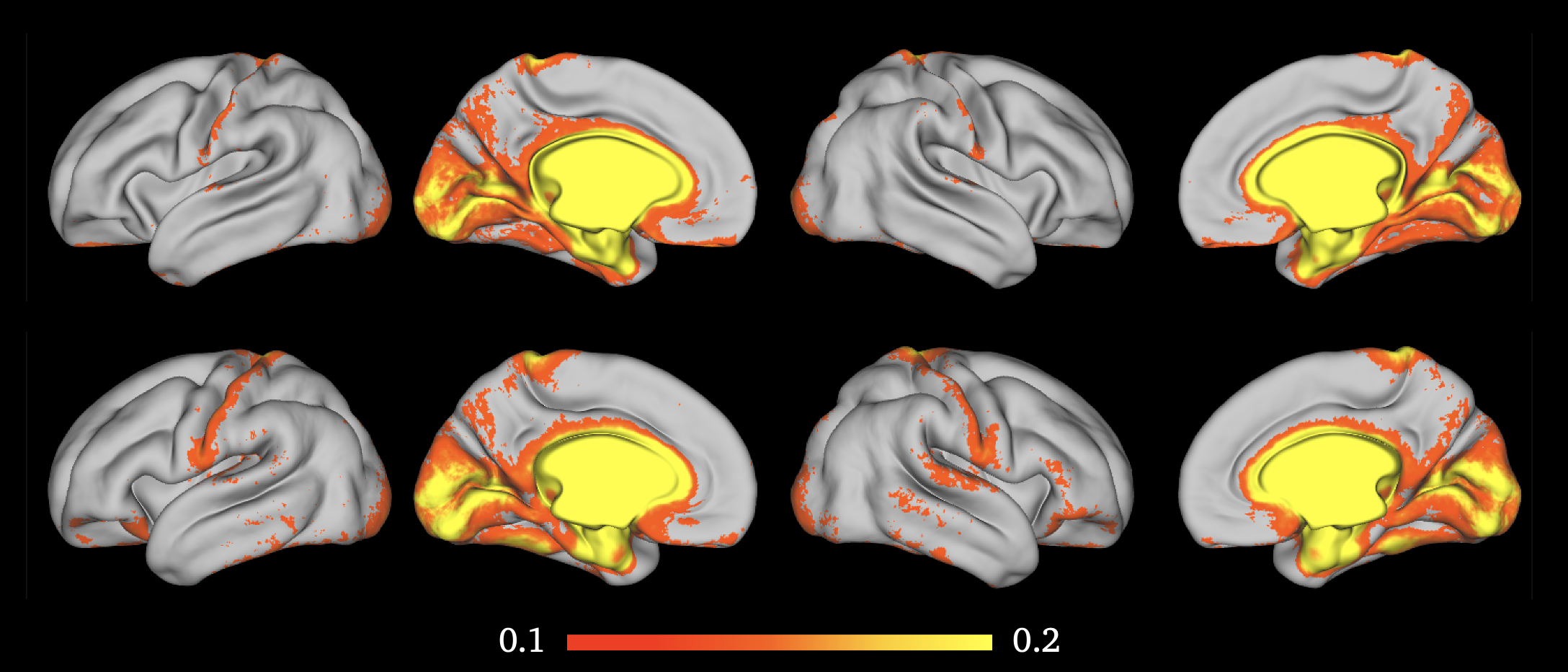}
\caption{The visualization of surface distances from the cortical surfaces reconstructed by our DL-based dHCP pipeline to those generated by the original dHCP pipeline. The averaged cortical surfaces are displayed and overlaid with surface distance maps ($0.1-0.2mm$) averaged over all test samples. Top: the averaged white matter surfaces and distance maps. Bottom: the averaged pial surfaces and distance maps.}
\label{fig:surf_distance}
\end{figure}

\begin{figure*}[ht]
\centering
\includegraphics[width=0.95\linewidth]{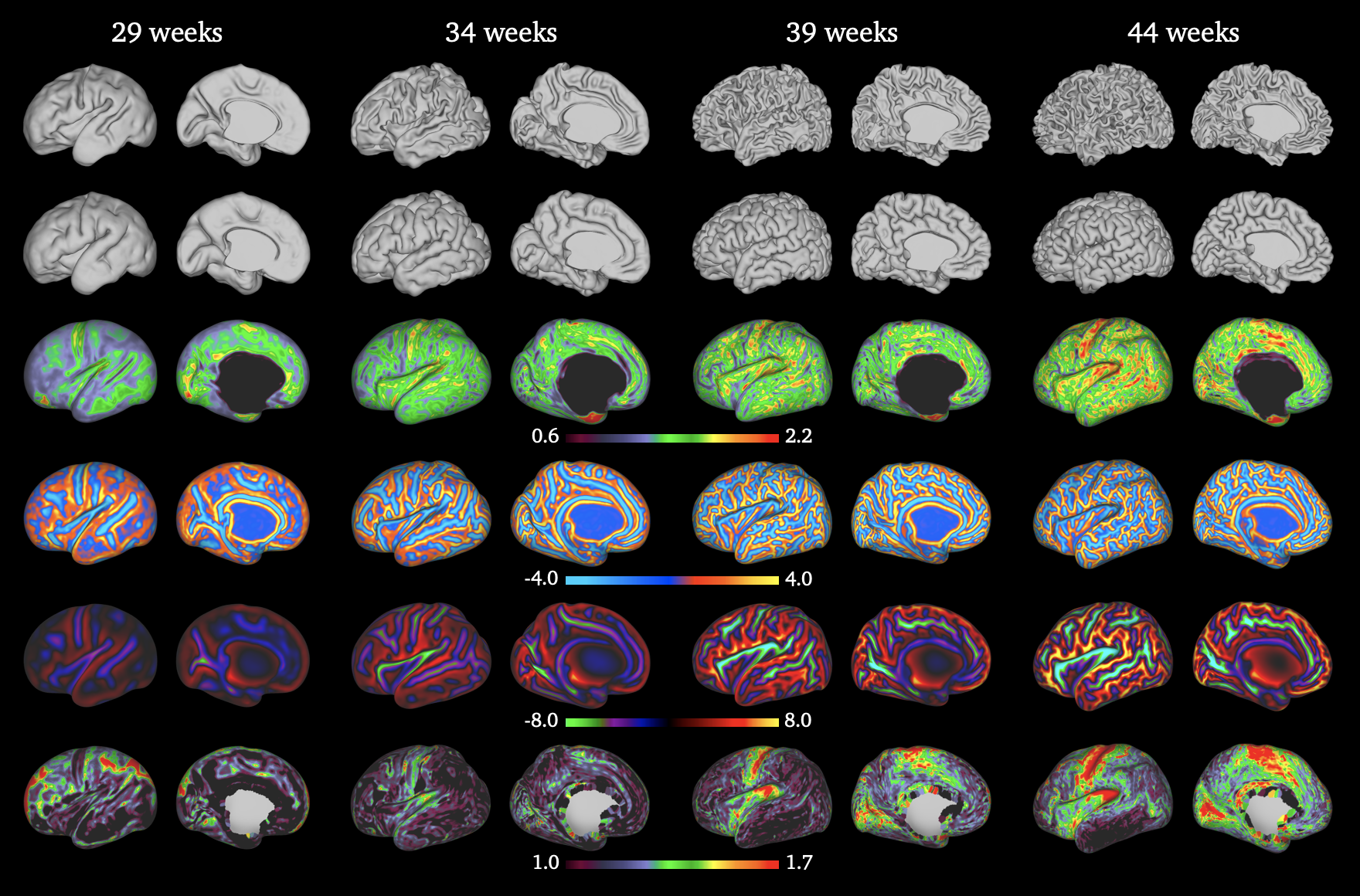}
\caption{The visualization of cortical surfaces and cortical features of different neonatal subjects at the PMA of 29,34,39,44 weeks. From top to bottom: white matter surface, pial surface, cortical thickness, mean curvature, sulcal depth, and myelin map.}
\label{fig:brain_develop}
\end{figure*}

\begin{figure*}[ht]
\centering
\includegraphics[width=0.49\linewidth]{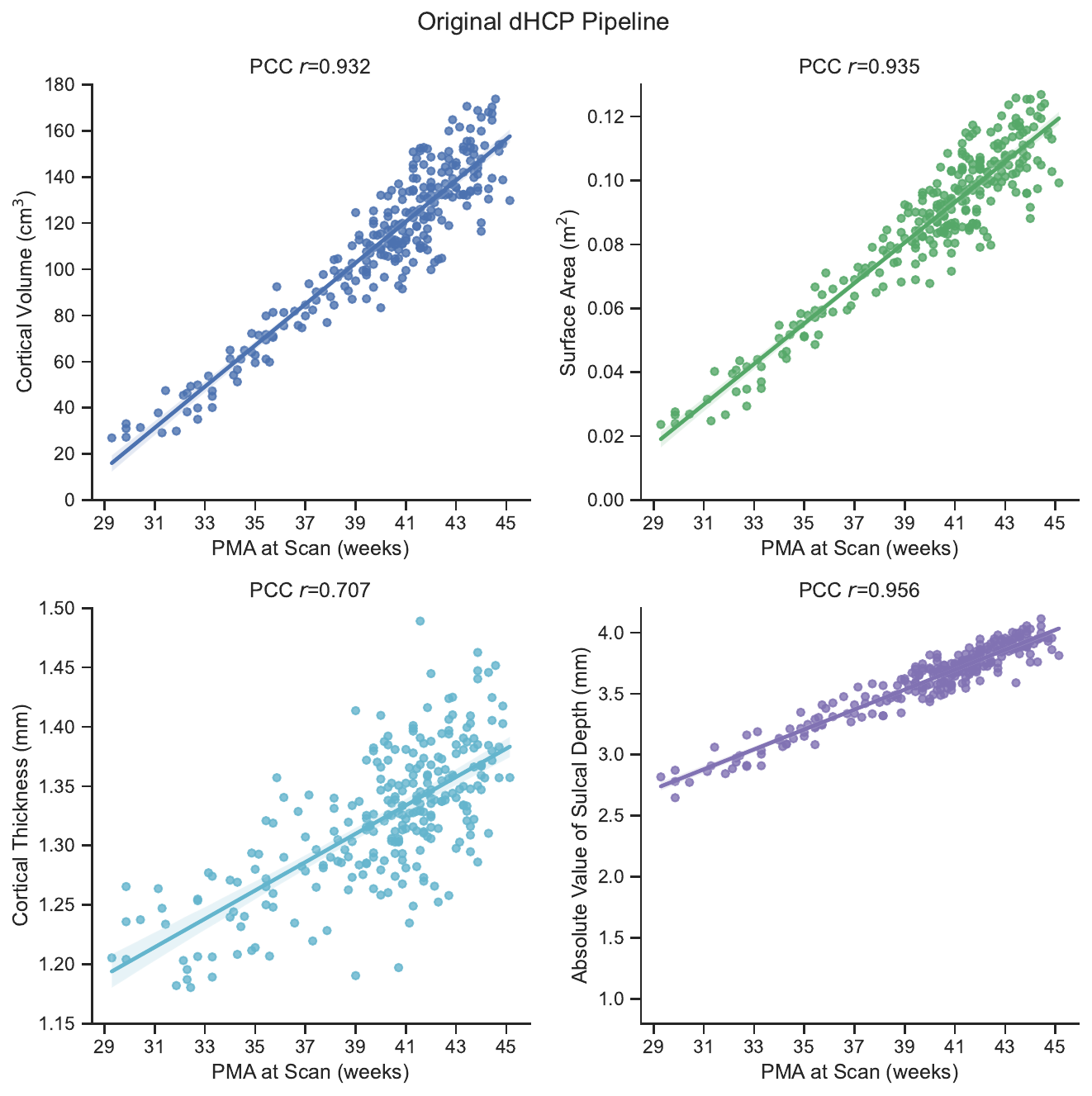}
\includegraphics[width=0.49\linewidth]{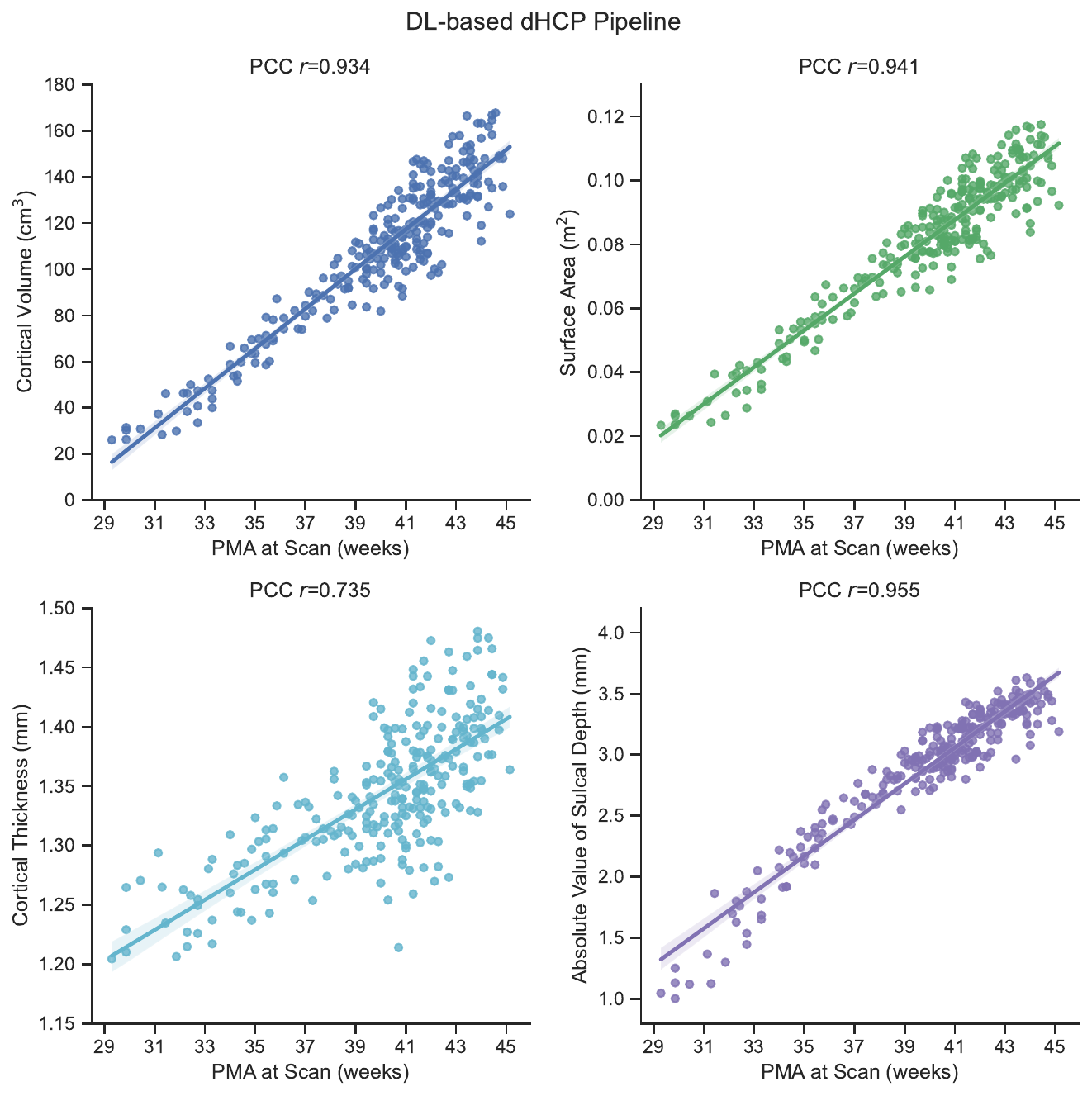}
\caption{Comparison on cortical morphological features including cortical volume ($cm^3$), surface area ($m^2$), average cortical thickness ($mm$), and average absolute value of sulcal depth ($mm$) across 265 test neonatal subjects with different PMA at scan. The Pearson correlation coefficients (PCC) $r$ are reported to measure the correlation between the morphological features and the PMA. Left: cortical morphological features estimated by the original dHCP pipeline. Right: cortical morphological features estimated by our DL-based dHCP pipeline.}
\label{fig:feat_stat}
\end{figure*}

\begin{figure*}[ht]
\centering
\includegraphics[width=0.94\linewidth]{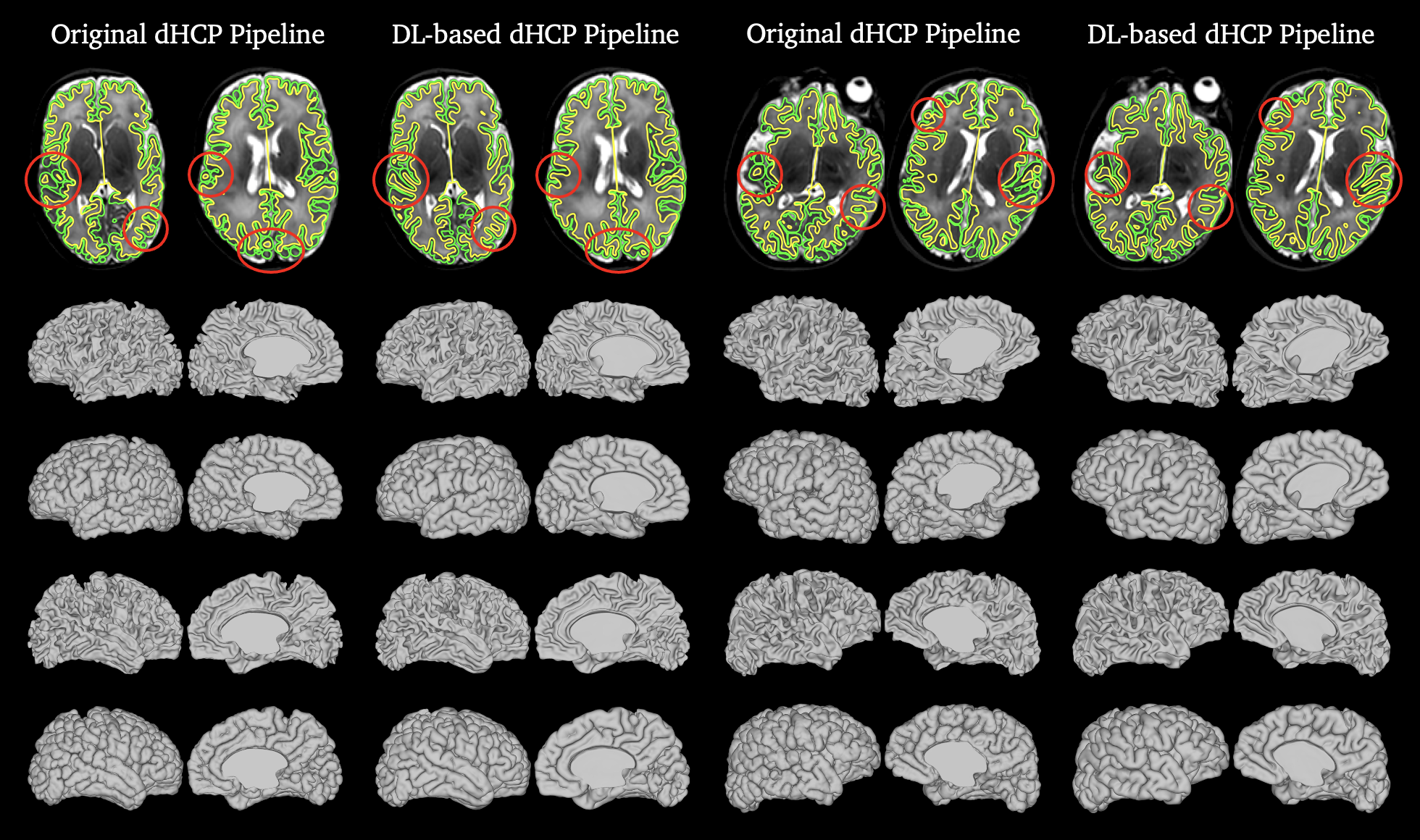}
\caption{The qualitative comparison between original and our DL-based dHCP pipeline on two neonatal subjects of the ePrime dataset. The cortical surfaces generated by the original dHCP pipeline are severely corrupted and broken on gyri, whereas our DL-based pipeline has better generalizability. From top to bottom: T2w MRI, left white matter surface, left pial surface, right white matter surface, and right pial surface. }
\label{fig:eprime}
\end{figure*}

\begin{figure*}[ht]
\centering
\includegraphics[width=0.86\linewidth]{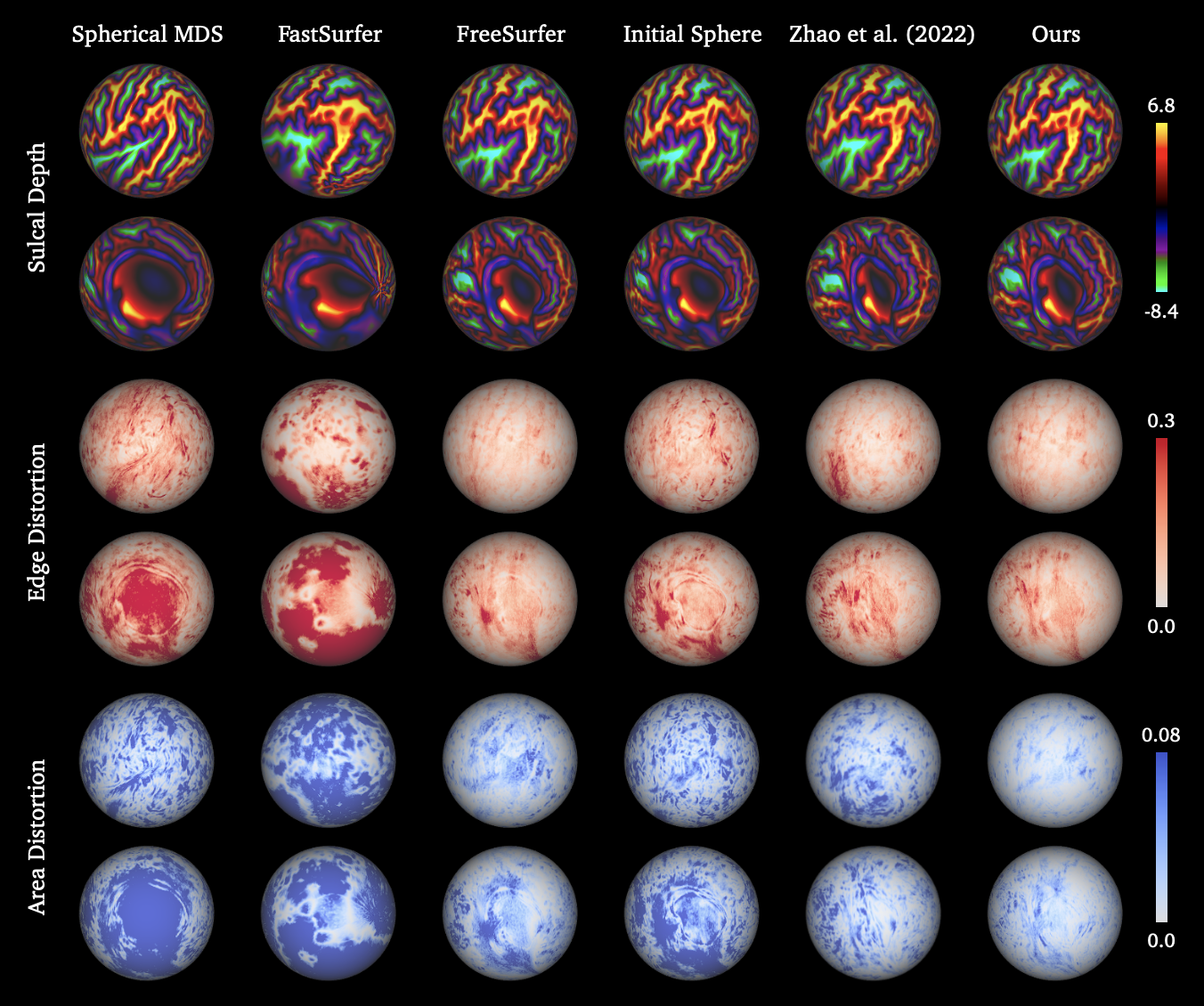}
\caption{The visualization of the projected spheres for all spherical mapping approaches. The projected spheres are overlaid with sulcal depth, edge and area distortions from top to bottom. For each vertex, the distortion is computed by averaging over the distortions of its adjacent edges or faces.}
\label{fig:sphere_quality}
\end{figure*}

\subsection{Comparison to DL-based Neuroimage Pipelines}\label{subsec:compare_infant}
We compare our dHCP DL pipeline to existing DL-based neuroimage processing pipelines, including FastSurfer \citep{henschel2020fastsurfer} and iBEAT V2.0 \citep{wang2023ibeat}. The comparisons regarding functionality and approximate runtime on a GPU are reported in Table \ref{tab:compare_learning}. Table \ref{tab:compare_learning} shows that previous DL-based pipelines focus on learning-based approaches for brain regional and tissue segmentations. However, theses pipelines still adopt traditional approaches for cortical surface reconstruction including computationally intensive topology correction \citep{bazin2005topology,bazin2007topology,segonne2007topology}, and thus require more than one hour to process a single brain MRI scan. Our DL-based dHCP pipeline introduces fast learning-based methods to accelerate the most time-consuming cortical surface reconstruction and spherical projection. Therefore, our pipeline is 150$\times$ and 600$\times$ faster than FastSurfer and iBEAT V2.0 respectively.

Our DL-based dHCP pipeline extracts cortical surfaces end-to-end from the T2w brain MRI without the need of intermediate tissue segmentations. This prevents the propagation of errors from segmentations to cortical surfaces. Therefore, as shown in Table \ref{tab:compare_learning}, currently our pipeline does not provide brain tissue segmentations and cortical surface parcellation. We will develop corresponding DL-based approaches and integrate them into our pipeline in future work.

\section{Quality Control}\label{sec:qc}

\subsection{Cortical Surface Quality}\label{subsec:qc-surf}

\subsubsection{Visual Inspection}
{In this section,} we conduct manual quality control (QC) via visual inspection to assess the quality of the cortical surfaces generated by our DL-based dHCP pipeline. 
{The cortical surface quality, \emph{i.e.}, the anatomical correctness of cortical surfaces, is measured by the number and severity of anatomical errors and corruptions in cortical surfaces. Due the lack of GT surfaces, we use the cortical surface generated by the original dHCP pipeline \citep{makropoulos2018dhcp} as the baseline.} More specifically, our visual QC aims to examine if our DL-based dHCP pipeline extracts cortical surfaces with better surface quality compared to the original dHCP pipeline.

We randomly select 30 T2w brain MRI samples for QC from 265 test cases. For each test brain MRI, we extract two groups of cortical surfaces by original dHCP pipeline and our DL-based dHCP pipeline respectively. The visual QC is performed by four expert assessors with clinical background or extensive experience in neuroimage analysis to mitigate individual variability. To ensure fairness, all assessors visually evaluate two groups of cortical surfaces independently without knowing the corresponding reconstruction approaches (either original or DL-based pipeline). The assessors are requested to carefully examine and detect corrupted areas in two groups of cortical surfaces, and then select the group with better surface quality for each test sample based on the number and severity of identified defects and artifacts. Equal quality is also an option when both groups of cortical surfaces are anatomically correct or corrupted to the same degree.

The average and individual QC results are reported in Figure~\ref{fig:qc}. The results show that for 82.5\% of all test samples, our DL-based dHCP pipeline achieves superior (54.2\%) or equal (28.3\%) surface quality compared to the original dHCP pipeline in visual assessment. The individual QC results in Figure \ref{fig:qc} also indicate that our DL-based dHCP pipeline achieves better performance in the qualitative evaluation of all assessors. These demonstrate that our DL-based dHCP pipeline reconstructs higher quality cortical surfaces, while being 995$\times$ faster than the original dHCP pipeline.

Additionally, we provide visual comparison between cortical surfaces reconstructed by original and DL-based dHCP pipeline. Figure \ref{fig:surf_quality} displays the cortical surfaces of 5 different samples. It indicates that the original dHCP pipeline is prone to produce corruptions on the top and back (parietal and occipital lobes) of the cortical surfaces, due to the low image contrast in the corresponding regions of T2w brain MRI, whereas our DL-based pipeline achieves better anatomical correctness in these regions compared to the original dHCP pipeline. In addition, the surfaces generated by original dHCP pipeline are affected by the imprecise Draw-EM segmentations \citep{makropoulos2014drawem,makropoulos2018dhcp}, and thus occasionally produce unexpected "caves" or "holes" as shown in Figure~\ref{fig:surf_quality}. Our DL-based dHCP pipeline leverages diffeomorphic surface deformations and learns cortical surfaces end-to-end from brain MRI, which effectively avoids such issues.

Based on the manual QC results, we further measure the visual QC score for each of 30 test samples. The visual QC score $s$ of each test subject is computed by $s=\sum_{i=1}^4 s_i$, where $s_i=1$ if the $i$-th assessor observes that our DL-based pipeline produces better surface quality compared to the baseline of the original dHCP pipeline, $s_i=-1$ if the original pipeline performs better, and $s_i=0$ means equal surface quality. For example, $s=4$ indicates all 4 assessors agree that our DL-based pipeline has better performance on surface quality. We count the number of test subjects grouped by the visual QC scores $-4\leq s\leq 4$. The results presented in Figure \ref{fig:qc_score} show that 80\% of the 30 test cases have positive visual QC scores $s>0$, which means our DL-based pipeline achieves better anatomical correctness than the original dHCP pipeline.

To explore both succeeded and failed cases of the proposed DL-based dHCP pipeline, in Figure \ref{fig:fail_case}, we visualize the test samples with QC scores $s=4$ and $s=-3$, \emph{i.e.}, all 4 assessors agree that the surface quality of our DL-based dHCP pipeline is better and no better than the original pipeline. Row 1-2 in Figure \ref{fig:fail_case} shows that the succeeded cases ($s=4$) can alleviate the systematic errors on the top and back of the cortical surfaces produced by the original dHCP pipeline. However, the failed examples ($s=-3$) of our DL-based pipeline in Row 3-4 still introduce corruptions in the occipital lobe, affected by the limited surface quality of the pseudo GT cortical surfaces extracted by the original dHCP pipeline.

\subsubsection{Quantitative Evaluation}
In addition to visual inspection, we conduct quantitative evaluation to validate that our DL-based dHCP pipeline can alleviate the anatomical errors produced by original dHCP pipeline \citep{makropoulos2018dhcp} in areas with low image contrast, \emph{e.g.}, the top and back (occipital lobe) of the brain. Following previous work \citep{zaretskaya2018distance,little2021distance,ma2022cortexode}, we compute the surface distances from the vertices of the cortical surfaces predicted by our DL-based pipeline to the pseudo GT cortical surfaces generated by the original pipeline. Since the cortical surfaces of our DL-based pipeline share the same mesh connectivity, we visualize the averaged cortical surfaces overlaid with the surface distance maps averaged over all 265 test cases in Figure \ref{fig:surf_distance}. 

The distance maps displayed in Figure \ref{fig:surf_distance} show large surface distances on the top, back (occipital lobe), and medial wall of the cortical surfaces. Combining with the manual QC results, the large quantitative differences on the top and back of the cortical surfaces imply that our DL-based pipeline can reduce the systematical errors introduced by original dHCP pipeline in these regions observed in Figure \ref{fig:surf_quality}. Besides, the large surface distance on the medial wall, which is out of the ROI in cortical surface analysis \citep{fischl1999cortical,glasser2013hcp,makropoulos2018dhcp}, is expected because original dHCP pipeline extracts cortical surfaces from segmentations that contain the clear boundary of the medial wall \citep{makropoulos2018dhcp}, whereas our DL-based pipeline extracts cortical surface end-to-end from brain MRI scans.

\subsubsection{Brain Development}

We provide qualitative and quantitative evaluation to validate that the cortical surfaces extracted by our dHCP DL pipeline can capture the brain development of neonates. We visualize the cortical surfaces as well as corresponding cortical features of different neonatal subjects scanned at the PMA of 29,34,39,44 weeks in Figure~\ref{fig:brain_develop}. The cortical folding is observed in Figure~\ref{fig:brain_develop} with the increasing of the PMA. The cortical thickness and sulcal depth are increasing as well with brain growth.

{For both original and DL-based dHCP pipeline}, we measure and plot cortical morphological features including cortical volume, surface area, cortical thickness and sulcal depth across all 265 test neonatal subjects in Figure~\ref{fig:feat_stat}. The cortical volume is equal to the volumes of pial surfaces minus those of the white matter surfaces for both brain hemispheres. The surface area refers to the mesh area of the pial surfaces. The cortical thickness and sulcal depth for each subject is measured by averaging the cortical thickness and the absolute value of the sulcal depth over all vertices on the surface meshes. 
Figure~\ref{fig:feat_stat} shows the positive trends of the morphological features with increasing PMA, which reflect the growth and folding of the neonatal cerebral cortex. Moreover, we compute the Pearson correlation coefficients (PCC) $r$ to measure the correlation between the morphological features and the PMA. The PCCs are provided in Figure~\ref{fig:feat_stat}, which demonstrate that all cortical morphological features have strong positive correlation with the age of the neonates. This verifies that the cortical surfaces extracted by our DL-based pipeline are able to capture the brain development of neonates as the original pipeline.

Compared to the original dHCP pipeline, the cortical volume and surface area measured by our DL-based pipeline manifest similar trends. For cortical thickness, our DL-based pipeline presents stronger correlation and higher PCC value than the original pipeline. We notice that in Figure~\ref{fig:feat_stat}, the absolute value of sulcal depth estimated by the original pipeline is much higher than ours. As mentioned in Section \ref{subsec:data-prepare} and \ref{subsec:surf-inflate}, the cortical surfaces extracted by original dHCP pipeline have adapted number of vertices, \emph{i.e.}, less vertices for younger subjects. As a consequence, the cortical surfaces of younger infants are over-inflated during the inflation process, thereby causing overestimation of sulcal depth. In contrast, our DL-based pipeline creates cortical surfaces with fixed number of vertices, ensuring age-consistent measurement of the sulcal depth across the PMA.

\begin{table}[ht]\small
\centering
\begin{tabular}{lcc}
\toprule
Method & White Matter Surface & Pial Surface \\ 
\midrule
Original dHCP pipeline & 0.457$\pm$7.409 & \textbf{0.517$\pm$7.418} \\
Forward Euler & 0.864$\pm$5.712 & 4.925$\pm$17.71\\
w/o Gaussian smoothing & 0.613$\pm$4.304 &  3.715$\pm$9.824 \\
Ours & \textbf{0.179$\pm$3.171} &  1.500$\pm$8.875 \\
\bottomrule
\end{tabular}
\caption{The number of self-intersecting faces (SIFs) on cortical surfaces.}
\label{tab:sif}
\end{table}

\subsubsection{Surface Topology}
Our DL-based dHCP neonatal pipeline is able to guarantee the spherical topology of cortical surfaces. The proposed pipeline uses multiscale deformation networks to learn diffeomorphic deformations to deform an initial surface to cortical surfaces. As the diffeomorphic deformations preserve the topology of the surfaces and the initial surface has the same topology as a sphere, the cortical surfaces are topologically equivalent to a sphere as well. The Euler characteristic number is 2 for all extracted surface meshes as expected.

We further examine surface self-intersections, which usually appear in the narrow sulci of pial surfaces and will affect the estimation of cortical morphological features \citep{ma2022cortexode}. We detect and count the number of self-intersecting faces (SIFs) in white matter and pial surface meshes generated by both original and DL-based dHCP pipelines. To verify that the proposed multiscale deformation networks can effectively reduce SIFs, we conduct ablation experiments by removing the Gaussian smoothing layer or replacing the scaling and squaring method \citep{higham2005scaling,arsigny2006scaling} with the forward Euler for integration. The number of SIFs are reported in Table~\ref{tab:sif}.

Although our multiscale deformation networks theoretically guarantee the diffeomorphisms, there are still negligible number of SIFs caused by numerical errors in the discretization, \emph{e.g.}, the ODE integration, the volumetric representation of deformation fields, and the limited number of mesh vertices. Table~\ref{tab:sif} shows that our DL-based dHCP pipeline produces only 0.18 SIFs on average in white matter surfaces. The original dHCP pipeline produces marginally fewer SIFs in pial surfaces than ours since it performs collision detection during the iterative surface deformation. Table~\ref{tab:sif} also validates that SIFs can be reduced effectively by adding the Gaussian smoothing layer and using the scaling and squaring method.

\subsubsection{Generalizability}
In addition to the dHCP neonatal dataset \citep{edwards2022release}, we evaluate the generalizability of our DL-based dHCP pipeline on an unseen dataset. We consider the Evaluation of Preterm Imaging (ePrime) dataset \citep{edwards2018eprime,grigorescu2021harmonized}, comprising 486 T2w MRI scans of neonates with gestational age between 23 and 33 weeks, in which 444 subjects are scanned at term-equivalent age between 38 and 45 weeks. The MR images are acquired with a Philips Intera 3T scanner, using a T2w turbo echo sequence (TES) with parameters repetition/echo time TR/TE=8670/160ms and TSE factor of 16. The ePrime dataset has relatively low image resolution with the voxel size of $0.86\times0.86\times1.0mm^3$, and thus it is challenging to extract high-quality cortical surfaces.

To test the generalizability, we run both original and our DL-based dHCP pipelines on unseen T2w MRI scans of the ePrime dataset without fine-tuning. The qualitative comparison in Figure \ref{fig:eprime} shows that there are severe corruptions and discontinuity on the gyri of cortical surfaces generated by the original dHCP pipeline, while our DL-based pipeline produces cortical surfaces with acceptable quality. This demonstrates that our DL-based pipeline has better generalizability on the unseen and low-resolution dataset.

\begin{table}[ht]\small
\centering
\begin{tabular}{lccc}
\toprule
Method & Edge ($mm$) & Area ($mm^2$) & Runtime\\ 
\midrule
Spherical MDS & 0.186$\pm$0.039 & 0.064$\pm$0.020 & 74min\\
FastSurfer & 0.284$\pm$0.064 & 0.106$\pm$0.033 & 2.73s\\
FreeSurfer & 0.123$\pm$0.025 & 0.033$\pm$0.010 & 4min12s\\
Initial Sphere & 0.165$\pm$0.033 & 0.055$\pm$0.017 & --\\
\cite{zhao2022sphereproj} & 0.139$\pm$0.026 & 0.035$\pm$0.010 & 2.11s\\
Ours & \textbf{0.119$\pm$0.023}$^*$ & \textbf{0.023$\pm$0.006}$^*$ & \textbf{0.33}s$^*$ \\
\bottomrule
\end{tabular}
\caption{Comparisons of metric distortions between spherical projection approaches. The edge distortion ($mm$), area distortion ($mm^2$) and runtime for each brain hemisphere are reported. $^*$Our learning-based approach produces significantly smaller distortions compared to all other approaches ($p<0.05$).}
\label{tab:distortion}
\end{table}

\subsection{Sphere Quality}\label{subsec:qc-sphere}

We evaluate the performance of the spherical projection by the metric distortions defined in Eq.~(\ref{eq:distortion}) on 265 test samples in the dHCP neonatal dataset. Similar to the original dHCP pipeline \citep{makropoulos2018dhcp}, the quality of the sphere is measured by the distortions of the edge length and face area between the white matter surface and the projected sphere. We compare our unsupervised learning-based approach with existing spherical mapping methods including Spherical MDS \citep{elad2005mds}, FreeSurfer \citep{fischl2012freesurfer}, FastSurfer \citep{henschel2020fastsurfer}, and \cite{zhao2022sphereproj}. We also provide the metric distortions between the white matter surface and the initial sphere as a baseline. 
For fair comparison, we only use a single-scale Spherical U-Net in \cite{zhao2022sphereproj} like our approach. The edge and area distortions, as well as the average runtime for each brain hemisphere are reported in Table~\ref{tab:distortion}. We conduct paired t-tests to examine if our unsupervised learning approach performs significantly better than other baseline methods.

Table~\ref{tab:distortion} shows that our approach achieves significantly better performance ($p<0.05$) compared to all baseline methods and only requires 0.33 seconds of runtime. As reported in Table~\ref{tab:distortion}, the runtime of Spherical MDS \citep{elad2005mds} is more than one hour due to the time-consuming pairwise geodesic distance computation. Despite the efficiency of the spherical embedding approach in FastSurfer \citep{henschel2020fastsurfer}, it does not optimize the metric distortions. FreeSurfer \citep{fischl2012freesurfer} achieves satisfactory performance but it takes 4 minutes to process each hemisphere. The initial sphere, which is generated by FreeSurfer as a good starting point, effectively reduces the baseline metric distortions without additional deformations.

Compared to \cite{zhao2022sphereproj}, our approach has several advantages. First, \cite{zhao2022sphereproj} needs to inflate and project the white matter surfaces to create initial spheres, which requires 1.79 seconds for each brain hemisphere with GPU acceleration in our experiments. Instead, our approach uses a fixed initial sphere for all subjects without extra processing. The initial sphere is generated by FreeSurfer with minimized metric distortions as reported in Table \ref{tab:distortion}. This provides a strong prior with small initial distortions for the training. Such a global initial surface also saves the runtime of interpolation and resampling, since the barycentric coordinates can be pre-computed. Hence, our approach is capable of projecting the cortical surface to a sphere in 0.33 seconds, which is 6$\times$ faster than \cite{zhao2022sphereproj}. Furthermore, \cite{zhao2022sphereproj} computes the metric distortions between the resampled white matter surface and the icosphere, and thus requires multiscale networks to improve the performance. As reported in Table \ref{tab:distortion}, \cite{zhao2022sphereproj} does not outperform FreeSurfer in the case of single-scale architecture. Our end-to-end metric distortion loss is more effective so that a single-scale Spherical U-Net is sufficient to produce significantly smaller distortions than FreeSurfer.

We also provide qualitative comparison as shown in Figure \ref{fig:sphere_quality}, which visualizes the projected spheres overlaid with cortical sulcal depth, edge and area distortions for all approaches. Figure \ref{fig:sphere_quality} shows that Spherical MDS and FastSurfer produce large distortions for edge length and face area. Compared to FreeSurfer, which is the best baseline method as reported in Table \ref{tab:distortion}, our unsupervised learning approach achieves similar visualization of sulcal depth, similar edge distortions, much smaller area distortions, and $760\times$ faster runtime.

\section{Discussion}\label{sec:discussion}
In this work, we presented a fast DL-based cortical surface reconstruction pipeline for structural brain MRI processing and analysis on the dHCP neonatal dataset. The proposed DL-based dHCP pipeline, which is primarily implemented in Python and PyTorch, consists of learning-based brain extraction, cortical surface reconstruction and spherical projection, as well as GPU-accelerated cortical surface inflation and feature estimation. The entire pipeline executes within only 24 seconds accelerated by GPU and takes less than 3 minutes even without GPU support. This is orders of magnitude faster than traditional neuroimage pipelines \citep{fischl2012freesurfer,glasser2013hcp,makropoulos2018dhcp} that require at least 6 hours of runtime, as well as existing DL-based cortical surface reconstruction pipelines \citep{henschel2020fastsurfer,wang2023ibeat} that need more than one hour.

The results of visual surface quality control validate that for more than 80\% of test samples, our DL-based dHCP pipeline produces cortical surfaces with higher (54.2\%) or equal (28.3\%) surface quality compared to the original dHCP pipeline \citep{makropoulos2018dhcp}. Particularly, the visual inspection on cortical surfaces and quantitative evaluation on surface distance demonstrate that our learning-based cortical surface reconstruction approach can effectively reduce the artifacts on the top and occipital lobe of the cortical surfaces produced by the original dHCP pipeline. We also verify that the proposed DL-based dHCP pipeline is capable of generalizing well on unseen and low-resolution ePrime neonatal brain MRI data \citep{edwards2018eprime, grigorescu2021harmonized}. In addition, our unsupervised learning-based spherical projection framework significantly reduces both edge and area distortions compared to previous work while only requires 0.33 seconds of runtime for each brain hemisphere.

However, despite the high efficacy and efficiency of the proposed dHCP DL pipeline, the anatomical correctness of predicted cortical surfaces is inevitably restricted by the surface quality of pseudo GT cortical surfaces generated by original dHCP pipeline, due to the limitations of supervised learning. Besides, the proposed multiscale deformation network predicts cortical surfaces from the input T2w brain MRI directly without the reliance on segmentations. Such an end-to-end framework facilitates the surface reconstruction procedure and avoids the potential errors in cortical surfaces introduced by imprecise segmentations, while this also means that our DL-based dHCP pipeline does not provide brain tissue segmentations and cortical surface parcellation.

In future work, we will consider semi-supervised learning technique to further enhance the anatomical correctness of the cortical surfaces. In addition to pseudo GT cortical surfaces, the unsupervised brain MRI intensity will be utilized to deform the cortical surfaces towards the WM/cGM or cGM/CSF boundary where the MRI achieves maximum image contrast. Furthermore, we plan to enrich the functionality and incorporate more GPU-accelerated processing steps into our DL-based dHCP pipeline, \emph{e.g.}, brain regional and tissue segmentations, cortical parcellation, and local gyrification index estimation.

\section*{Acknowledgments}
Qiang Ma is funded by the President’s PhD Scholarship at Imperial College London. Kaili Liang is supported by National Institute for Health Research (NIHR) Maudsley Biomedical Research Centre (BRC) PhD studentship. Liu Li is supported by Lee Family Scholarship. Support was also received from the ERC project MIA-NORMAL 101083647 and ERC project Deep4MI 884622. The dHCP neonatal dataset was provided by the developing Human Connectome Project, KCL-Imperial-Oxford Consortium funded by the ERC under the European Union Seventh Framework Programme (FP/2007-2013) / ERC Grant Agreement no. [319456]. We are grateful to the families who generously supported this trial.

% Acknowledgments should be inserted at the end of the paper, before the
% references, not as a footnote to the title. Use the unnumbered
% Acknowledgements Head style for the Acknowledgments heading.

% \section*{References}

% Please ensure that every reference cited in the text is also present in
% the reference list (and vice versa).

%%Harvard
\bibliographystyle{model2-names.bst}\biboptions{authoryear}
\bibliography{ref}

% \section*{Supplementary Material}

% Supplementary material that may be helpful in the review process should
% be prepared and provided as a separate electronic file. That file can
% then be transformed into PDF format and submitted along with the
% manuscript and graphic files to the appropriate editorial office.

\end{document}